\documentclass[twocolumn,pre,showpacs,amsmath,amssymb]{revtex4}
\usepackage{color,epsfig,bm,amsmath,amssymb,hyperref}
\newcommand{\cN}{{\cal N}}

\newcommand{\bbbN}      {{\mathbb{N}}}                  % natural numbers
\newcommand{\bbbsone}   {{\mathrm{1\hspace{-0.7mm}I}}}  % small unit matrix

\newcommand{\rd} {\mathrm d}
\newcommand{\re} {\mathrm e}

\newcommand{\vk} {{\bm k}}

\newcommand{\vx} {{\bm x}}

\newcommand{\nn}{\nonumber} 
\newcommand{\be}{\begin{equation}} 
\newcommand{\ee}{\end{equation}}  
\newcommand{\bea}{\begin{eqnarray}}
\newcommand{\eea}{\end{eqnarray}}
\newcommand{\barr}{\begin{array}}
\newcommand{\earr}{\end{array}}
\newcommand{\bcent}{\begin{center}} 
\newcommand{\ecent}{\end{center}}  

\begin{document}
%%%%%%%%%%%%%%%%%%%%%%%%%%%%%%%%%%%%%%%%%%%%%%%%%%%%%%%%%%
\title{The Fate of Articulation Points and Bredges in Percolation}
\author{Haggai Bonneau$^1$, Ido Tishby$^1$, Ofer Biham$^1$, Eytan Katzav$^1$, and Reimer K\"uhn$^2$}
\affiliation{$^1$Racah Institute of Physics, The Hebrew University, Jerusalem, 9190401, Israel\\
$^2$Mathematics Department, King's College London, Strand, London WC2R 2LS, United Kingdom}
\date{\today}
\begin{abstract}
\noindent
We investigate the statistics of articulation points and bredges (bridge-edges) in complex networks in which bonds are randomly removed in a percolation process. Articulation points are nodes in a network which, if removed,  would split the network component on which they are located into two or more separate components, while bredges are edges whose removal would split the network component on which they are located into two separate components. Both articulation points and bredges play an important role in processes of network dismantling and it is therefore useful to know the evolution of the probability of nodes or edges to be articulation points and bredges, respectively, when a fraction of edges  is randomly removed from the network in a percolation process. Due to the heterogeneity of the network, the probability of a node to be an articulation point, or the probability of an edge to be a bredge will not be homogeneous across the network. We therefore analyze full distributions of articulation point probabilities as well as bredge probabilities, using a message-passing or cavity approach to the problem, as well as a deconvolution of these distributions according to degrees of the node or the degrees of both adjacent nodes in the case of bredges. Our methods allow us to obtain these distributions both for large single instances of networks as well as for ensembles of networks in the configuration model class in the thermodynamic limit of infinite system size. We also derive closed form expressions for the large mean degree limit of Erd\H{o}s-R\'enyi networks.
\end{abstract}

\maketitle
%%%%%%%%%%%%%%%%%%%%%%%%%%%%%%%%%%%%%%%%%%%%%%%%%%%%%%%%%%
\section{Introduction}
%%%%%%%%%%%%%%%%%%%%%%%%%%%%%%%%%%%%%%%%%%%%%%%%%%%%%%%%%%
Networks affect many dimensions of human existence. They manifest themselves in everyday life, they underpin advanced information and communication technologies, and they provide a powerful paradigm to analyse complex problems in the natural sciences, in engineering, and in economics and the social sciences \cite{HavlCoh2010, NewBk10, Estrada2011,BarrBarthVes08,Latora+2017}. Key functionality of a network often depends of the fact that pairs of nodes are mutually connected through one or several paths of contiguous edges, or whether on the contrary they reside in different disconnected components of a net. One of the key questions of network science has therefore been the identification of conditions under which networks --- natural or artificial --- exhibit a so-called giant connected component (GCC), which occupies a finite fraction of a system in the limit of large system size \cite{Bollobas1998}. A natural question then is how random or intentional removal of nodes or edges will affect the functionality of a network, in particular whether a GCC would survive such a process of node or edge removals; this question has been extensively studied in the past two decades using percolation theory  \cite{Albert2000, Cohen+2000, Call+00, Cohen+2001, Schneider+2001, HavlCoh2010, NewBk10, Estrada2011,BarrBarthVes08,Latora+2017}.

\begin{figure}[h!]
\includegraphics[width=0.42\textwidth]{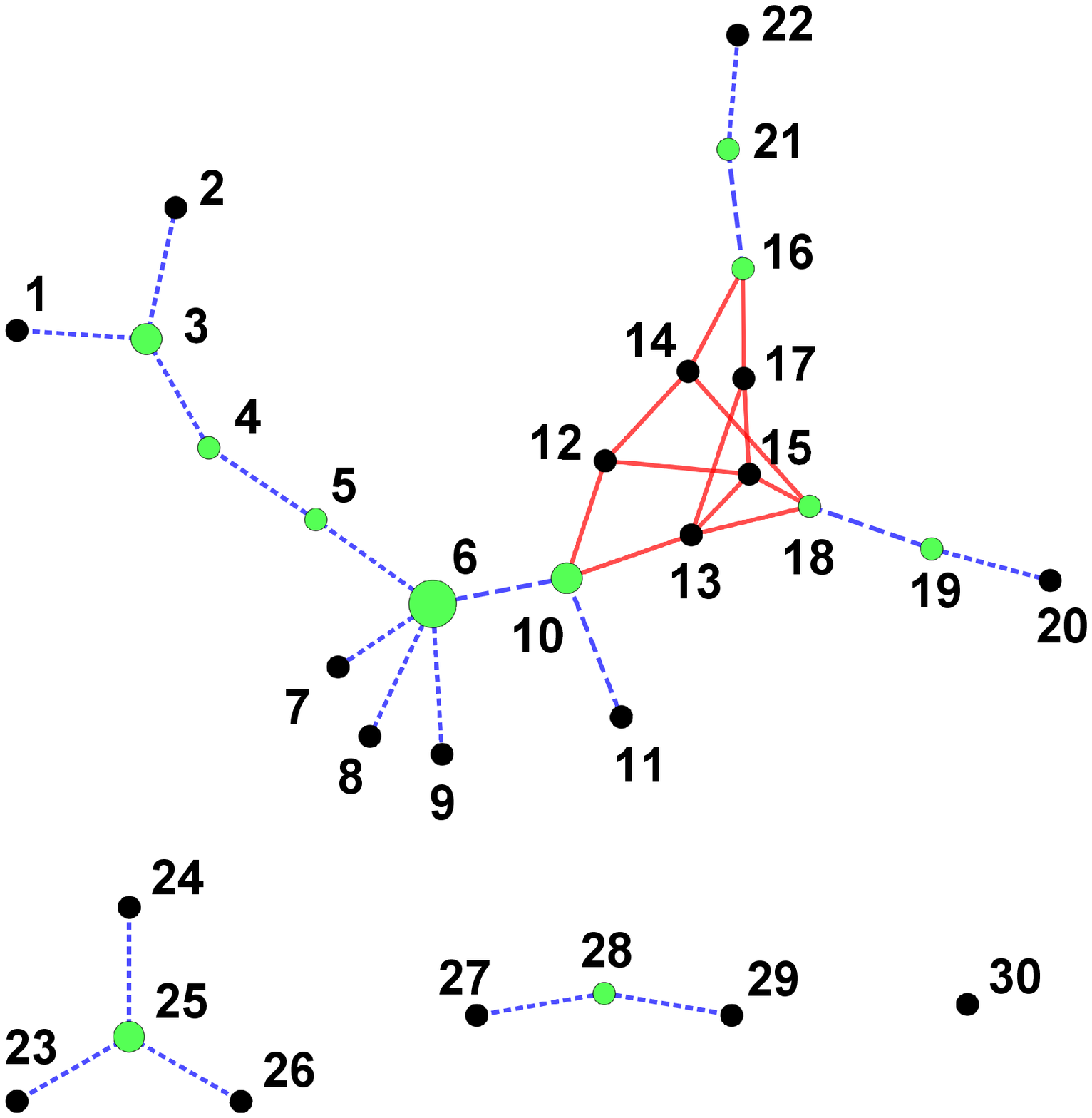}
\caption{(Color online) Articulation points and bredges in an ER network of mean degree $c=2$ and $N=30$ nodes. articulation points are indicated by circles, with a radius proportional to their articulation rank, i.e. proportional to the number of new network components that would be created by their removal, while vertices that are not articulation points are indicated as black dots. Bredges are indicated by broken lines, with long dashed lines indicating so-called root bredges which are directly linked to the 2-core of the network, and short dashes indicating the remaining bredges located on tree branches of the GCC or on finite isolated clusters. Full lines indicate edges which are part of at least one cycle, and hence their removal would not break the network into two components. These non-bredge edges form the so-called 2-core of a network.}
\label{fig:AP-bredgeExample}
\end{figure}

Of crucial importance for the functionality of a networks are articulation points (APs) and bredges (or bridge-edges) \cite{Bredge1685}. Articulation points are {\em nodes\/} whose removal would break the network component on which they are located into two or more disconnected components \cite{Hopcroft1973, Chaudhuri1998, Tian+2017, Wu+18}, while  bredges are {\em edges\/} whose removal would break the network component on which they are located into two components \cite{Tarjan1974, Italiano2012}. APs and bredges thus play a central role in network attack strategies where network dismantling is achieved by systematic removal of cycles, a process known as decycling  \cite{Braunstein+PNAS16, Zdeborova+2016, Wandelt+SciRep18}, which in turn generates further APs and bredges. It is clear from their definition that all nodes on trees that are not be leaf-nodes of a network are APs, and conversely that nodes that belong to any cycle cannot be APs, unless they are also root nodes of a tree. In a similar vein, any edge located on a subtree of a network will always be a bredge. Figure \ref{fig:AP-bredgeExample} provides an illustrative example.

\begin{figure}[t!]
\setlength{\unitlength}{1mm}
\begin{picture}(77,100)
\put(0,55){\includegraphics[width=0.42\textwidth]{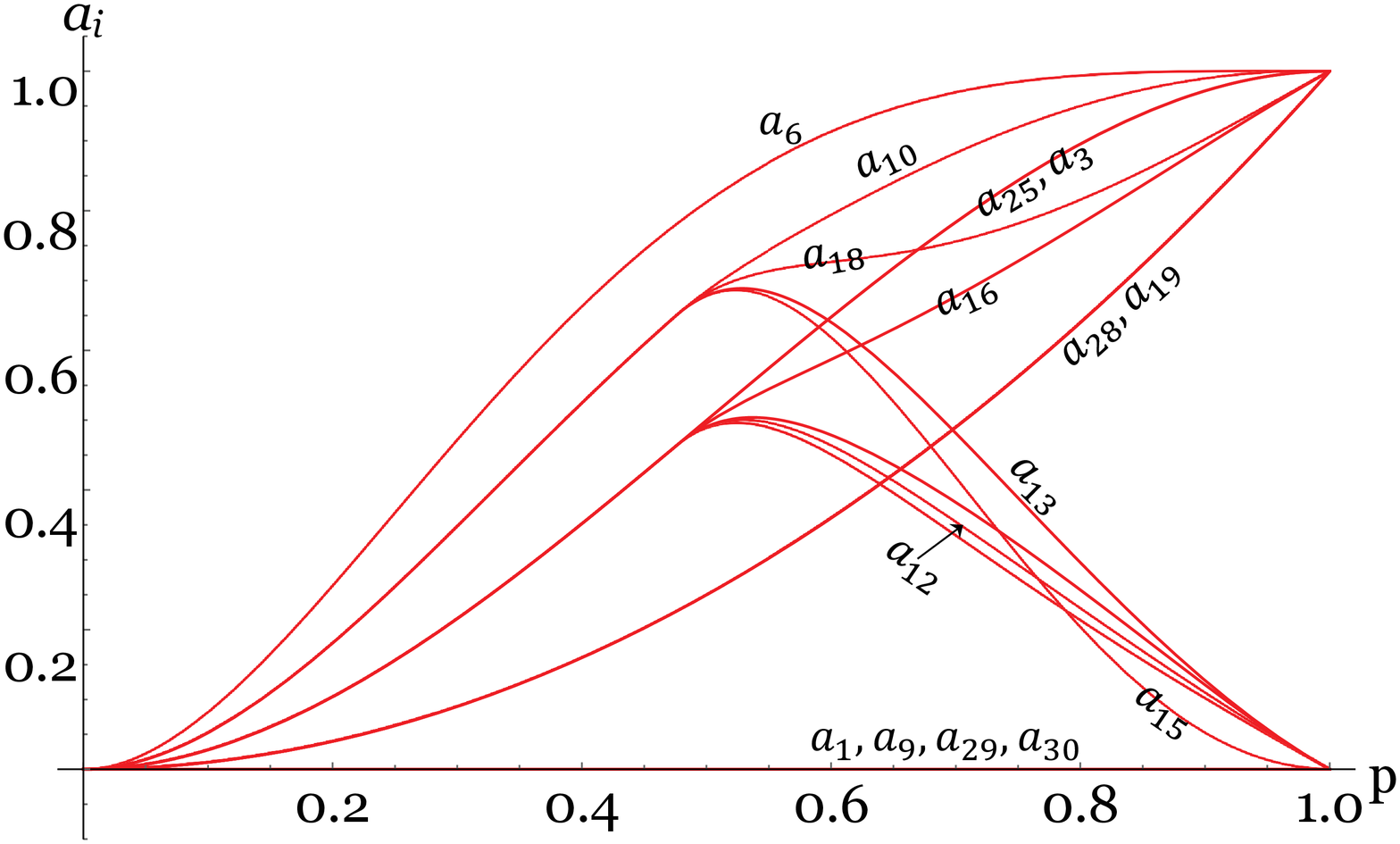}}
\put(0,0){\includegraphics[width=0.42\textwidth]{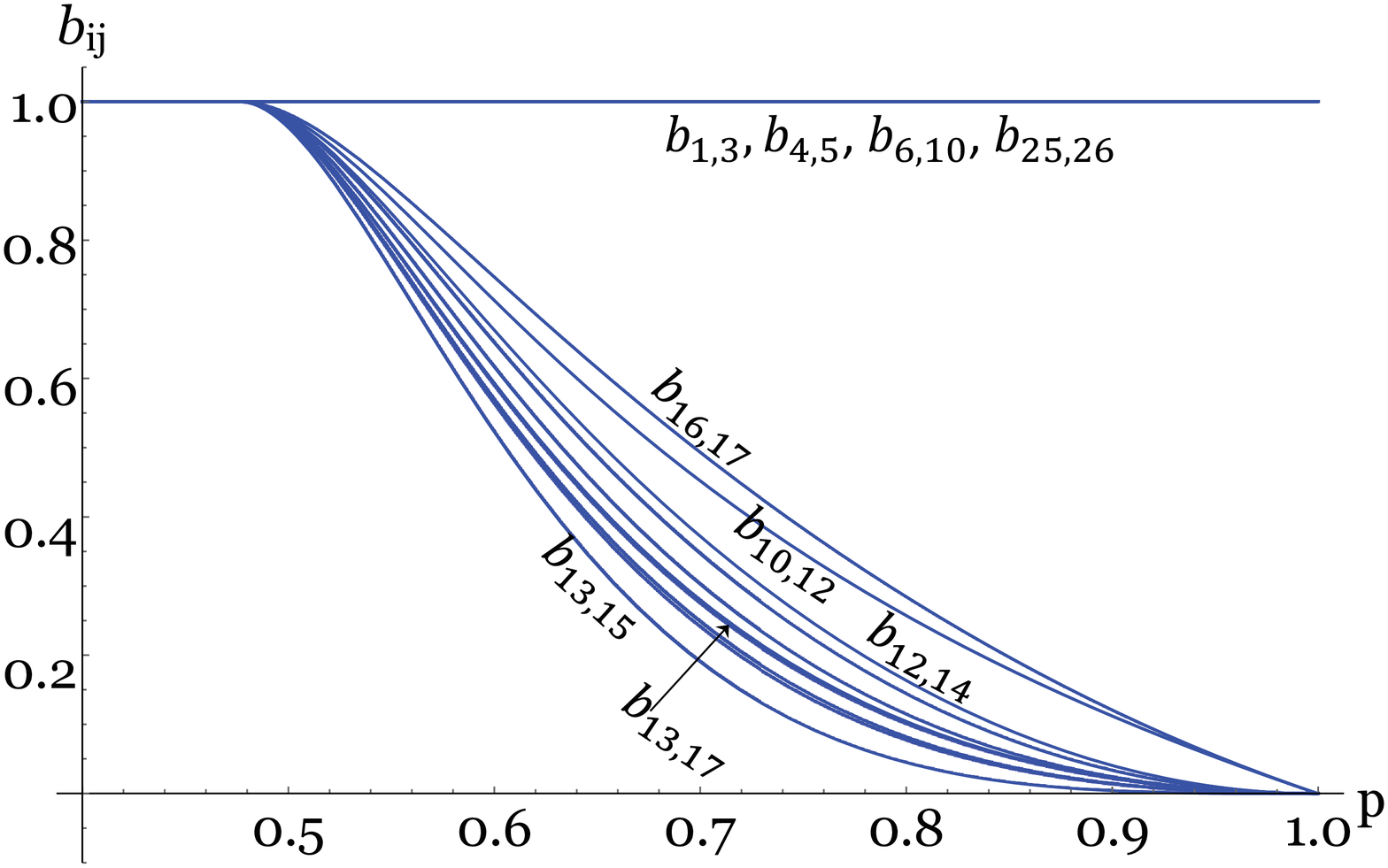}}
\put(-7,42){\large{\bf (b)}}
\put(-7,97){\large{\bf (a)}}
\end{picture}
\caption{(Color online) {\bf (a)}: Articulation point probabilities $a_i$ of a selected set of nodes $i$, and {\bf (b)}: bredge probabilities $b_{ij}$ of a selected set of edges $(i,j)$ from the network of Fig.\,\ref{fig:AP-bredgeExample} as functions of the bond retention probability $p$ in a percolation process.}
\label{fig:AP-bredgeProbs}
\end{figure}

It is worth mentioning, however, that there is also a constructive role for APs and bredges, as these are precisely the nodes or edges in a network one needs to remove in efficient strategies for containment (islanding) of blackouts in power grids, for containment of financial shocks or vaccination in the context of epidemics to name but a few relevant cases. From a different perspective, APs and bredges are fundamental quantities in describing the geometry of networks.

The statistical properties of APs  and bredges were recently studied in detail for models in the configuration model class in \cite{Tishby+18b} and \cite{Bonneau+20}, respectively. Closed form expressions were obtained for the average fraction of nodes that are APs, both in the entire network, and on the GCC of configuration model networks. Further details concerning the degree distribution of APs as well as the distribution of articulation ranks which specify the number of additional network components created by the removal of a node were also evaluated in closed form; for further details see  \cite{Tishby+18b}. In \cite{Bonneau+20}, analogous results were obtained for bredges, including the probability for edges to be bredges, both in the entire network, and separately for the GCC, as well as joint degree distributions of nodes connected by a bredge, once more both in the entire network, and restricted to the GCC. Both sets of results heavily rely on an earlier analysis of the micro-structure of the GCC and the finite network components of configuration models in \cite{Tishby+18}. Figure \ref{fig:AP-bredgeExample} shows an example of  an Erd\H{o}s-R\'enyi (ER) network exhibiting APs and bredges.

In \cite{Tishby+18b} and \cite{Bonneau+20}  the authors looked at average AP and bredge probabilities and their deconvolution according to degree for ensembles of random networks in the configuration model class. In the present paper we look at the evolution of these probabilities in a percolation process where a certain fraction of edges is randomly removed from the network with probability $1-p$ (and hence retained with probability $p$).  Moreover, we go beyond {\em average\/} probabilities and their deconvolution according to degree in recognition of the fact that the probability of a node to be an AP or of an edge to be a bredge will depend on higher order coordination shells around the node or edge in question, rather than just on the degree of a node or the two degrees of the end-nodes of an edge. This type of heterogeneity was first properly highlighted and analyzed in detail for percolation probabilities and cluster-size distributions in \cite{KuRog17}, and we will use a variant of that analysis for the study of APs and bredges.

Figure \ref{fig:AP-bredgeProbs} illustrates the heterogeneity of AP and bredge probabilities that is to be expected in this problem. It  shows AP probabilities for a selected set of  of nodes and bredge probabilities  for a selected set of edges for the example network of Fig.\,\ref{fig:AP-bredgeExample} as functions of the bond retention probability $p$ of a percolation process. Results are obtained using the theory outlined in the present paper, as predicted by Eq. \eqref{eq:ai} for APs and Eq.\,\eqref{bij} for bredges. They demonstrate, in particular, that nodes on the 2-core of the network at $p=1$ cannot be APs, {\em but may become APs\/} as $p$ is decreased and cycles are are eliminated through bond removal. Conversely nodes on tree-components of the network that are not endpoints at $p=1$ are APs with probability 1, but may lose that property by becoming endpoints or isolated nodes through bond removal as $p$ is decreased. For bredges, the figure illustrates in a similar vein that edges that belong to tree components of the network, whether attached to the GCC or not, are bredges with probability 1, while edges located on cycles on the original network, {\em i.e.} at $p=1$, are bredges with probability 0 {\em but may become bredges\/} as cycles are broken with increasing probability as $p$ is decreased. Our paper explains how this fairly intricate phenomenology can be captured analytically.

Our main results are the following. We demonstrate that the message passing approach to percolation probabilities can be used to evaluate node dependent probabilities of vertices in a complex networks to be APs as well as edge dependent probabilities of pairs of neighboring vertices in a network to be connected by a bredge. We derive a formulation for single instances of large networks and use that to obtain a formulation for ensembles of networks in the configuration model class in the thermodynamic limit. We obtain closed form approximations for the large mean degree limit of Erd\H{o}s-R\'enyi (ER) networks which we find to efficiently capture probability density functions of AP and bredge probabilities already for relatively moderate mean degrees. Distributions of AP probabilities and bredge probabilities are evaluated for ER networks as well as scale free networks in the thermodynamic limit, and we also obtain deconvolutions of these distributions according to degree(s) of  the node(s) involved. We also apply the single instance theory to obtain distributions of AP and bredge probabilities for a real-world network. Finally, we use the message passing approach to identify APs and bredges in a given network, and for any given realization of a percolation process, and find that it performs far better than might be expected, given that the method is known to be exact only on trees.

The remainder of this paper is organized as follows. In Sect.\,\ref{sec:Perc} we introduce the version of the message passing approach to percolation that we are using for the analysis of APs and bredges. We formulate the approach both for the analysis of finite instances of large networks, and for the thermodynamic limit of networks in the configuration model class. In Sect.\,\ref{sec:APBredges} we use these results to analyze the probabilities of edges to be bredges and the probabilities of nodes to be APs as a function of the edge retention probability $p$. Once more we  do this for large single network instances and for networks in the configuration model class, when the thermodynamic limit of infinite system size is taken. In Sect.\,\ref{sec:LargeMeanC} we derive closed form approximations for the probability density functions (pdfs) of AP and bredge probabilities for the large mean degree limit of Erd\H{o}s-R\'enyi (ER) networks. Results are presented in Sect.\,\ref{sec:Res}. Section \ref{sec:SumDisc} finally concludes with a summary and a discussion that also considers various possible applications of our main results.

%%%%%%%%%%%%%%%%%%%%%%%%%%%%%%%%%%%%%%%%%%%%%%%%%%%%%%%%%%
\section{Bond Percolation}
\label{sec:Perc}
%%%%%%%%%%%%%%%%%%%%%%%%%%%%%%%%%%%%%%%%%%%%%%%%%%%%%%%%%
The analysis of AP and bredge probabilities is closely related to the analysis of percolation in complex networks, a process by which bonds (or vertices) of a network are randomly and independently either kept with probability $p$ or deleted with probability $1-p$. The present investigation is based on a message passing approach to percolation in complex networks \cite{ShirKaba10, Karrer+14}. We will, however, complement the results of these papers by evaluating the results of the message passing approach {\em beyond\/} average percolation probabilities or, for that matter, average AP and bredge probabilities, using ideas proposed in \cite{KuRog17}. The approach taken in \cite{KuRog17} evaluates distributions of percolation probabilities in terms of size distributions of finite clusters. Here we use a more direct approach closer to \cite{ShirKaba10} which is formulated directly in terms of node dependent percolation probabilities.

We will formulate the message passing approach for single instances of large complex networks, and then use the results to obtain a description for networks in the configuration model class in the thermodynamic limit. Networks in the configuration model class are maximally random subject to a prescribed degree distribution \cite{Annibale+09, CooAnnibRob17}. Using $k_i$ to denote the degree of node $i$ in a network we thus assume that a network is characterized by a degree distribution ${\rm Prob}(k_i = k) = p_k$ for $k\in\bbbN$, and that there are no degree-degree correlations.

%%%%%%%%%%%%%%%%%%%%%%%%%%%%%%%%%%%%%%%%%%%%%%%%%%%%%%%%%%
\subsection{Single-Instance Theory for Percolation Probabilities}
%%%%%%%%%%%%%%%%%%%%%%%%%%%%%%%%%%%%%%%%%%%%%%%%%%%%%%%%%%
We begin by briefly describing the message passing approach to percolation, concentrating for specificity on {\em bond  percolation\/}. The case of site percolation where nodes are randomly removed with probability $1-p$ and retained with probability $p$ can be analyzed using the same ideas and methods.
 
We consider networks consisting of $N$ vertices, labeled $i=1, 2, \dots,N$ which are connected by a set of non-directed edges $(ij)$. We introduce indicator variables $n_i$ to denote whether vertex $i$ is in the giant connected component (GCC) of the network ($n_i=1$) or not ($n_i=0$), and indicator variables $x_{ij}$ which denote whether the edge $(ij)$ is kept in a single realization of the percolation process ($x_{ij}=1$) or not ($x_{ij}=0$). In terms of these we have 
\be
n_i = 1 - \prod_{j\in\partial i} \Big(1 -x_{ij} n_j^{(i)}\Big)
\label{ni-eq}
\ee
in which $\partial i$ denotes the set of nodes connected to $i$ in the original graph, and $n_j^{(i)}$ is an indicator variable denoting whether the vertex $j$ adjacent to $i$ is ($n_j^{(i)}=1$) or is not ($n_j^{(i)}=0$) on the GCC on the cavity graph from which vertex $i$ and the edges emanating from it are removed. Equation \eqref{ni-eq} states the fact that a site belongs to the GCC if it is connected to it through at least one of its neighbors, which in turn requires that both that neighbor is in the GCC, {\em and\/} that the link connecting to that neighbor is actually kept in the given instance of a percolation experiment. For the cavity indicator variables we have, by the same line of reasoning, that
\be
n_j^{(i)} = 1 - \prod_{\ell\in\partial j\setminus i}\Big(1 -x_{j\ell} n_\ell^{(j)}\Big)\ .
\label{nji-eq}
\ee
Equations (\ref{ni-eq}) and (\ref{nji-eq}) can be averaged over all possible realizations of the percolation process on the given network. This gives
\be
g_i = 1 - \prod_{j\in\partial i}\Big(1 - p g_j^{(i)}\Big)
\label{avni-eq}
\ee
for the probability that vertex $i$ will be part of the GCC in a realization of the percolation
process, while
\be
g_j^{(i)} = 1 - \prod_{\ell\in\partial j\setminus i}\Big(1 -p g_\ell^{(j)}\Big)
\label{avnji-eq}
\ee
is the probability that vertex $j$ neighboring on $i$ on the original graph will be part of the GCC on the cavity graph from which $i$ and edges emanating from it are removed. In Eqs.\, \eqref{avni-eq} we exploited the fact that that $x_{ij}$ and $n_j^{(i)}$ are independent, and the same clearly holds for $x_{j\ell}$ and $n_\ell^{(j)}$ in Eq.\,\eqref{avnji-eq}. As usual, a factorization of averages  that assumes independence of random variables along different edges incident on a  given node. This assumption is exact only on trees, but is known to be an excellent approximation on locally tree-like graphs, which becomes exact for finitely coordinated systems in the thermodynamic limit $N\to \infty$ of infinite system size.

Eqs.\,(\ref{avnji-eq}) can be solved through forward iteration --- starting from random initial conditions --- on a single instance of a graph, and from the solution site dependent percolation probabilities $g_i$ can be computed using Eqs.\,\eqref{avni-eq}. Alternatively, for certain random network ensembles, distributions of percolation probabilities can be evaluated in the thermodynamic limit.
%%%%%%%%%%%%%%%%%%%%%%%%%%%%%%%%%%%%%%%%%%%%%%%%%%%%%%%%%%
\subsection{Thermodynamic Limit}
%%%%%%%%%%%%%%%%%%%%%%%%%%%%%%%%%%%%%%%%%%%%%%%%%%%%%%%%%%
We will evaluate distributions of percolation probabilities, and subsequently distributions of articulation point probabilities and distributions of bredge probabilities in the thermodynamic limit for networks in the configuration model class. 

In the thermodynamic limit Eqs.\,(\ref{avnji-eq}) constitute an infinite system of coupled self-consistency equations for the cavity probabilities $g_j^{(i)}$. Assuming that a statistical law or a probability density $\tilde \pi(\tilde g)$ of the $g_j^{(i)}$ exists, it can be found, following meanwhile standard arguments  \cite{MezPar01, Rog+08, Ku15, KuRog17} by demanding probabilistic self-consistency. The value of $\tilde \pi(\tilde g)$ is obtained by summing probabilities of all instances of of the r.h.s. of Eqs.\,(\ref{avnji-eq}) for which $g_j^{(i)}\in (\tilde g,\tilde g+ \rd\tilde g]$. Using this procedure, Eqs.\,(\ref{avnji-eq}) result in 
\be
\tilde \pi(\tilde g) = \sum_k\! \frac{k}{c}\,p_k\!\! \int\!\! \Big[\prod_{\nu=1}^{k-1} \rd \tilde 
\pi(\tilde g_\nu)\Big] \delta\Big(\tilde g - \Big[1-\prod_{\nu=1}^{k-1} (1-p \tilde g_\nu)\Big]\Big)\ ,
\label{pit-gt}
\ee
in which $ \frac{k}{c}\,p_k$ is the probability that a randomly chosen neighbour of a node has degree $k$, and we have adopted the shorthand $\rd \tilde \pi(\tilde g_\nu) =  \tilde\pi(\tilde g_\nu)\, \rd  \tilde g_\nu$. Although this equation is a highly non-linear integral equation, it can be solved efficiently and to any desired degree of precision (limited only by computational power) using a population dynamics algorithm \cite{MezPar01}. In terms of the solution of Eq.\,(\ref{pit-gt}), the distribution $\pi(g)$ of node dependent percolation probabilities is found from Eq.\,(\ref{avni-eq}) as
\be
\pi(g) = \sum_k p_k \int \Big[\prod_{\nu=1}^{k} \rd \tilde \pi(\tilde g_\nu)\Big]
\delta\Big(g - \Big[1-\prod_{\nu=1}^{k} (1-p \tilde g_\nu)\Big]\Big)\ .
\label{pi-g}
\ee

%%%%%%%%%%%%%%%%%%%%%%%%%%%%%%%%%%%%%%%%%%%%%%%%%%%%%%%%%%
\section{Statistics of Articulation Points and Bredges in Percolation}
\label{sec:APBredges}
%%%%%%%%%%%%%%%%%%%%%%%%%%%%%%%%%%%%%%%%%%%%%%%%%%%%%%%%%%
%%%%%%%%%%%%%%%%%%%%%%%%%%%%%%%%%%%%%%%%%%%%%%%%%%%%%%%%%%
\subsection{Articulation Points}
\label{sec:APs}
%%%%%%%%%%%%%%%%%%%%%%%%%%%%%%%%%%%%%%%%%%%%%%%%%%%%%%%%%%
In order for a node $i$ of the system {\em not\/} to be an articulation point,  all  its neighbors must reside on the giant component of the reduced network from which i is removed \cite{Tishby+18b}. Introducing $\hat n_i \in\{0,1\}$ as an indicator variable which denotes whether $i$ is an articulation point ($\hat n_i =1$) or not ($\hat n_i =0$), and noting that {\em only the links that are still present}, for which thus $x_{ij}=1$, should contribute to the logic as to whether or not a node is an articulation point, we get
\bea
\hat n_i &=& \Big[1 - \prod_{j\in\partial i} \big (n_j^{(i)}\big)^{x_{ij}}\Big] \times \bbbsone_{|\vx_{\partial i}| \ge 2}\nn\\
&=& \Big[1 - \prod_{j\in\partial i} \big(1-x_{ij} + x_{ij} n_j^{(i)}\big)\Big]\!\! \times\!\!\bbbsone_{|\vx_{\partial i}| \ge 2} \ .
\label{hatni}
\eea
Here we have introduced the vector $\vx_{\partial i} = (x_{ij})_{j\in\partial i}$ and the norm $|\vx_{\partial i}| = \sum_{j\in\partial i} x_{ij}$, and we have invested the fact that nodes connected to fewer than 2 other nodes cannot be articulation points. Upon averaging this over realizations of a percolation process, this gives the probability 
\be
a_i=  \langle \hat n_i \rangle
\ee
that node $i$ is an articulation point, where angled brackets denote an average over bond configurations in an ensemble of percolation processes in which bonds (of a given network) are randomly and independently removed with probability $1-p$ and kept with probability $p$. Performing the average over bond configurations in Eq.\,\eqref{hatni} we get
\bea
 a_i  &=& \bigg\langle\Big[1 - \prod_{j\in\partial i} \big(1-x_{ij} + x_{ij} n_j^{(i)}\big)\Big]\!\! \times\!\!\bbbsone_{|\vx_{\partial i}| \ge 2}\bigg\rangle\nn\\
& =& 1- p (1-p)^{k_i-1} \sum_{j\in\partial i}\big (1- g_j^{(i)}\big)  \nn\\
& &\hspace{20mm} -  \prod_{j\in\partial i} \big(1-p + p g_j^{(i)}\big)\ .
\label{eq:ai}
\eea
The heterogeneity of the original network entails that the $a_i$ depend in a highly non-trivial way on the location of the nodes $i$ in the original network. Following the reasoning used above to obtain the distribution of percolation probabilities, one obtains the probability density function $\pi(a)$ of the node dependent articulation point probabilities in the thermodynamic limit of infinite system size as
\be
\pi(a)  = \big(p_0+p_1\big)\, \delta(a)  + \sum_{k\ge 2} p_k  \pi(a|k)
\label{pi-a}
\ee
with
\bea
\pi(a|k) &=& \int\!\!  \Big[\prod_{\nu=1}^{k} \rd  \tilde\pi(\tilde g_\nu) \Big] \delta\bigg(a- \bigg[1 -  p (1-p)^{k-1}\nn\\
& & \times  \sum_{\nu=1}^k\big(1-  \tilde g_\nu \big) -\prod_{\nu=1}^k \big(1-p + p \tilde g_\nu\big)\bigg]\bigg)
\label{pi-ak}
\eea
giving the pdfs of articulation point probabilities conditioned on degrees for which $k\ge 2$.

Using the fact that  $0 \le g_j^{(i)}\le 1$ for all $g_j^{(i)}$ appearing in Eq.\,\eqref{eq:ai}, it is straightforward to obtain $p$-dependent expressions for  articulation point probabilities of and limiting probabilities for some subclasses of vertices. For example, for any node $i$ residing on a finite cluster of a network --- examples are nodes labeled 25, 28, 29 an 30 in Fig.\,\ref{fig:AP-bredgeExample} --- we have $g_j^{(i)}=0$ for all $j\in\partial i$.  For these nodes Eq.\,\eqref{eq:ai} then entails that
\be
a_i\big|_{k_i=k} = a_k^{\rm FC}(p) = 1 - k p (1-p)^{k-1}  - (1-p)^k\ .
\label{AFCkp}
\ee
These form a family of continuous curves for which $a_k^{\rm FC}(0)=0$  and $a_k^{\rm FC}(1)=1$, and the $a_k^{\rm FC}(p)$ would mark $p$-dependent locations of $\delta$-peaks in pdfs of AP probabilities for any network in which finite isolated clusters exist. Curves labeled $a_{25}$, $a_{28}$, $a_{28}$ and 
$a_{30}$ in Fig.\,\ref{fig:AP-bredgeProbs} provide examples belonging to this family. Given that isolated clusters are {\em generated\/} through percolation whenever $p <1$ these curves should eventually become noticeable in pdfs of AP probabilities even in systems in which finite clusters to not exist at $p=1$. Below the percolation threshold, they completely describe the support of the distribution of AP probabilities.

Next, suppose that $i$ is a node on the GCC of a network with $k_i=k$, and suppose that $k_t < k-1$ of the edges emanating from $i$ belong to a {\em tree\/} rooted in $i$, whereas the remaining $k_\ell = k-k_t\ge 2$ edges emanating from $i$ are part of one or several loops on the GCC. Examples of such nodes are nodes 10, 16, and 18 in Fig.\,\ref{fig:AP-bredgeExample}. Their corresponding $p$ dependent AP probabilities $a_i(p)$ as given by Eq.\,\eqref{eq:ai} are marked in Fig.\,\ref{fig:AP-bredgeProbs}. For nodes of this type we have $g_{j_t}^{(i)}=0$ for all $j_t\in \partial i$ which are located on the tree, whereas for the the remaining $k_\ell$ neighbors of $i$, have the the inequality $(1-g_{j_\ell}^{(i)}) \le (1-p)^{k_{j_\ell}-1}$ which follows by using the upper bound 1 in $g_\ell^{(j_\ell)} \le 1$ for all $\ell \in\partial j_\ell\setminus i$ in Eq.\,\eqref{avnji-eq}.  Note that $g_\ell^{(j_\ell)}$'s close to (the upper bound ) 1 are only likely to  be found sufficiently far above any percolation transition, thus for $p \lesssim 1$. Denoting by $\vk=(k_{j_\ell})$ the set of  degrees of the $k_\ell$ terminal nodes that link $i$ to loops, we can conclude that for vertices of this type we have
$$
a_i\big|_{k_i=k;k_t,\vk} \to  a_{k;k_t,\vk}(p)
$$
with
\bea
a_{k;k_t,\vk}(p) & = & 1 -  p (1-p)^{k-1}\Big[ k_t + \sum_{\ell=1}^ {k_\ell}(1-p)^{k_{j_\ell}-1}\Big]\nn\\
& & - (1 - p)^{k_t} \prod_{\ell=1}^{k_\ell} \big[1 - p (1-p)^{k_{j_\ell}-1} \big]\ .
\label{AGCkp}
\eea
These form families of curves  for which  $a_{k;k_t,\vk}(0) =0$, just as in the family of curves that describe the situation on finite clusters. If $k_t>0$ (and $i\in$ GCC), i.e. if $i$ is a root-node of a tree attached to the GCC, (examples are nodes 10, 16, and 18 in Fig.\,\ref{fig:AP-bredgeExample}) , then $a_{k;k_t,\vk}(p) \to  1$ as $p\to 1$, just as in the case of finite clusters.  However, if $k_t=0$ (and $i\in$ GCC), then $i$ is {\em not\/} the root node of a tree attached to the GCC (examples are nodes 12, 13, 14, 15, and 17 in Fig.\,\ref{fig:AP-bredgeExample}), and we have  $a_{k;k_t,\vk}(p) \to 0$ as $p\to 1$ {\em unlike\/} in the finite cluster case. Close to $p=1$ it is expected that the probability of  $g_\ell^{(j_\ell)}$'s saturating their upper bound is expected to be reasonably high, so this family of curves is expected to be reasonably well visible, as they would correspond to locations of pronounced maxima in $p$-dependent pdfs of AP probabilities, at least in networks which are reasonably densely connected at $p=1$. We shall find that this is clearly borne out by the results presented below.

In order to rationalize further structures in $p$-dependent pdfs of AP probabilities, one would have to include information about the configuration of higher coordination shells around a chosen vertex $i$, and use iterated versions of the self-consistency equation \eqref{avnji-eq} to express the  $g_{j}^{(i)}$ for $j\in \partial i$ in terms of cavity percolation probabilities on edges further removed from $i$. Following that strategy, one would in principle be able to characterize the $p$ dependence of such structures in terms of sums of powers of $p$ and $(1-p)$. In that context, the small example provided in Figs.\,\ref{fig:AP-bredgeExample} and \ref{fig:AP-bredgeProbs} can be instructive.

%%%%%%%%%%%%%%%%%%%%%%%%%%%%%%%%%%%%%%%%%%%%%%%%%%%%%%%%%%
\subsection{Bredges}
\label{sec:Bs}
%%%%%%%%%%%%%%%%%%%%%%%%%%%%%%%%%%%%%%%%%%%%%%%%%%%%%%%%%%
Moving on to bredges, we can follow the same line of reasoning. For a randomly chosen edge $(ij)$ in a network {\em not\/} to be a bredge \cite{Wu+18, Bonneau+20}, both of its end-nodes must belong to the GCC in a network from which the edge $(ij)$ is removed. Introducing $n_{ij}$ as an indicator variable that denotes whether the edge $(ij)$  is a bredge ($n_{ij}=1$) or not ($n_{ij}=0$), this can be expressed in terms of the cavity indicator variables $n_j^{(i)}$ and $n_i^{(j)}$ introduced above as
\be
n_{ij} = 1 - n_i^{(j)} n_j^{(i)}
\label{nij}
\ee
Averaging this equation over all realizations of the percolation process gives the probability
\be
b_{ij} = \langle n_{ij}\rangle = 1 - g_i^{(j)} g_j^{(i)}\ ,
\label{bij}
\ee
for a link $(ij)$ to be a bredge in an ensemble of percolation processes where links (on a given network) are randomly and independently removed with probability $1-p$ and kept with probability $p$. Equation \eqref{bij} allows one to obtain link-dependent bredge-probabilities $b_{ij}$ in large single network instances from the solutions of Eqs.\,\eqref{avnji-eq}. Once again, the heterogeneity of the original network entails that the $b_{ij}$ depend in a non-trivial way on the location of the edge $(ij)$ in the original network. The probability density function $\pi(b)$ of the link dependent bredge-probabilities in the thermodynamic limit is obtained following the reasoning used to find the distribution of percolation probabilities as
\be
\pi(b) = \int \rd\tilde\pi(\tilde g) \rd \tilde \pi(\tilde g')\,\delta\big(b - (1-\tilde g \tilde g')\big) \ .
\label{piofb}
\ee

To access the dependence of bredge probabilities on the degrees of the terminal nodes of an edge, one can use the self-consistency equations \eqref{avnji-eq} in Eq.\,\eqref{bij} giving
\be
b_{ij} = 1 - \Big [1- \!\!  \prod_{\ell\in\partial i\setminus j}\!\!\big(1 -p g_\ell^{(i)}\big)\Big]
\Big[ 1-  \!\! \prod_{{\ell'}\in\partial j\setminus i}\!\!\big(1 -p g_{\ell'}^{(j)}\big)\Big]
\label{bij-it}
\ee
In the thermodynamic limit this then translates into
\be
\pi(b) =  \sum_{k,k'} \frac{k}{c}\,p_k \frac{k'}{c}\,p_{k' \,} \pi(b|k,k')\ ,
\label{pi-b}
\ee
with the
\begin{widetext}
\bea
\pi(b|k,k') &=& \int \Big[\prod_{\nu=1}^{k-1} \rd \tilde \pi(\tilde g_\nu)\Big] \Big[ \prod_{\nu'=1}^{k'-1} \rd \tilde \pi(\tilde g_{\nu'})\Big] \,
%\Big]\nn\\& & \times 
\delta\Bigg(b -\bigg\{1- \Big[1-\prod_{\nu=1}^{k-1} (1-p \tilde g_\nu)\Big]
%\nn\\& & \hspace{13mm}\times 
 \Big[1-\prod_{\nu'=1}^{k'-1} (1-p \tilde g_{\nu'})\Big]\bigg\}\Bigg)
\label{pi-bkkprime}
\eea
\end{widetext}
as pdfs of the bredge probabilities, conditioned on the degrees $k$ and $k'$ of the terminal nodes of an edge.

For bredge probabilities described by Eq.\,\eqref{bij} one can obtain $p$-dependent families of curves that depend on the local environment of the terminal nodes $i$ and $j$ defining the edge. If $i$ or $j$ (or both) are nodes residing on a tree of the original network then the product $g_i^{(j)} g_j^{(i)}$ is identically zero, hence for edges of this type we have
\be
b_{ij} = b_{ij}^t(p) \equiv 1\ .
\ee
Examples are edges (1,3), (4,5), (6,10) and (25,26) in Fig.\,\ref{fig:AP-bredgeExample}, with the corresponding $b_{ij}(p)$ curves marked in Fig.\,\ref{fig:AP-bredgeProbs} (b).
This results in the appearance of a $\delta$-peak at $b=1$ in the pdf of the bredge probabilities at all $p$ (provided the network does contain trees (be they attached to the GCC or not). If networks are constructed without tree components at $p=1$, i.e., prior to random bond dilution, then this $\delta$-peak at $b=1$ will appear with increasing weight as $p$ decreases.

Assuming that an edge $(ij)$ connects nodes with $k_i=k$ and $k_j=k'$, and that both are indeed on the GCC of a network from which the edge in question is removed, then one can use the bounds $g_i^{(j)} \le 1 - (1-p)^{k_i-1}$ and similarly $g_j^{(i)} \le   1 - (1-p)^{k_j-1}$ to conclude
\be
b_{ij}\big|_{k_i=k,k_j=k'} \ge  b_{k,k'}(p)
\ee
with
\be
 b_{k,k'}(p) = 1 -\big [(1 - (1-p)^{k-1}\big] \big [(1 - (1-p)^{k'-1}\big]
\label{BGCkkpp}
\ee
As discussed above in the case of AP probabilities one expects that close to $p=1$ the probability of $g_i^{(j)}$'s and $g_j^{(i)}$'s saturating their upper bound should be reasonably high so this family of curves is expected to be reasonably well visible in representations of $p$-dependent pdfs of bredge probabilities, as long as the networks are fairly densely connected in the $p\to 1$-limit.

As in the case of AP probabilities further structures in the results for bredge probabilities can be rationalized by considering higher order coordination shells around the two vertices defining an edge under study.

%%%%%%%%%%%%%%%%%%%%%%%%%%%%%%%%%%%%%%%%%%%%%%%%%%%%%%%%%%
\section{Large Mean Degree Approximation}
\label{sec:LargeMeanC}
%%%%%%%%%%%%%%%%%%%%%%%%%%%%%%%%%%%%%%%%%%%%%%%%%%%%%%%%%%
For networks exhibiting `narrow' degree distributions in the sense that the the standard deviation of the degrees is negligibly  small in comparison to the mean degree, it is possible to derive closed form approximations of the results above  \cite{KuRog17, KuvM20}. The obvious candidate to consider is the Poisson degree distribution of ER graphs with large mean degree $\langle k\rangle = c$, for which the standard deviation $\sigma_k=\sqrt{c}$ is small compared to the mean for $c\gg 1$.

In the large mean degree limit the solution of Eq.\,\eqref{pit-gt} turns out to be well approximated by the $\delta$-distribution
\be
\tilde\pi(\tilde g) = \delta(\tilde g - \tilde g_*)\ .
\label{eq:deltaAnsatz}
\ee
The value of $g_*$ is obtained by inserting this ansatz into Eq.\,\eqref{pit-gt}, and deriving a self-consistency equation for $g_*$. Assuming a Poisson distribution for the degrees, we get the equation
\be
 g_* = 1 - \re^{-p c g_*}\ .
\label{gstar}
\ee
as the self-consistency equation for $g^*$. In order to obtain a non-trivial solution in the large $c$ limit, one has to adopt the scaling $p = \rho/c$ at fixed $\rho$, so that Eq.\,\eqref{gstar} becomes
\be
 g_* = 1 - \re^{-\rho g_*}\ ,
\label{gstarn}
\ee
which can be solved in closed form, giving
\be
g_* =  1 + \frac{W(-\rho \re^{-\rho})}{\rho}\ ,
\ee
where $W(\cdot)$ is the Lambert $W$-function; see Sect.\,4.13 in \cite{NIST:DLMF}.

In order to obtain the large mean degree limit of the distribution $\pi(a)$ of articulation point probabilities, we insert the ansatz of Eq.\,\eqref{eq:deltaAnsatz} into Eq.\,\eqref{pi-ak}, which implies that the conditional probability $a(k)$ for a node of degree $k$ to be an articulation point is at large $k (\ge 2)$ given by
\be
a(k) = 1 - k p (1-p)^{k-1}(1 - g_*)- (1-p+p g_*)^k
\label{aofk}
\ee
For a Poisson distribution of mean degree $c$, the distribution of scaled degrees $x = k/c$ is well approximated by a normal distribution of mean 1 and variance $1/c$ for $c\gg 1$, i.e., $x \sim \cN(1,1/c)$ in this limit. From Eq.\,\eqref{aofk}, we can obtain an expression of the scaled degree $x=x(a)$ as a function of the AP probability $a$ as the solution $x=x(a)$ of
\bea
a &=& 1 - \rho x (1-p)^{cx-1}(1 - g_*)- (1-p+p g_*)^{cx}\nn\\
  &\simeq& 1 -  \rho x \,\re^{-\rho x}(1 - g_*) - \re^{-\rho x (1-g_*)}\ ,
%x =x(a) = \frac{\ln(1 - a)}{c \ln(1-p+p g_*)} \ .
\label{xofa}
\eea
where we have used the large $c$ limit in the second line. This then allows us to obtain a closed form expression for the pdf $\pi(a)$ using the fact that the distribution of scaled degrees $x$ is a normal distribution
\be
\pi(x) = \sqrt{\frac{c}{2\pi}}\,\exp\Big[- \frac{c}{2} (x-1)^2\Big]
\label{eq:piofx}
\ee
which, via a standard identity about transformations of pdfs under a change of variable, transforms into
\bea
\pi(a) &=&\pi(x)\Big|\frac{\rd x}{\rd a}\Big| \nn\\
&=& \frac{ \sqrt{\frac{c}{2\pi}} \exp\Big[-\frac{c}{2}
\left(x -1\right)^2\Big]}{\big|\rho(1-g_*) \re^{-\rho x}\big[(\rho x-1)+\re^{\rho x g_*}\big]\big|}\ ,	
\label{piofg-largec}
\eea
with $x=x(a)$ given in terms of the solution of Eq.\,\eqref{xofa}. Note that the most efficient way to evalutate this density, however, is to avoid solving Eq.\,\eqref{xofa}, but to simply treat the {\em pair of equations} \eqref{xofa} and \eqref{piofg-largec} as a {\em parametric representation\/} of the pdf $\pi(a)$ in terms of the parameter $x \ge 0$.

The large mean degree approximation for the pdf $\pi(b)$ of bredge probabilities is slightly more involved, as according to Eqs.\,\eqref{pi-b} and \eqref{pi-bkkprime} it involves a natural deconvolution on contributions depending on the degrees of both end-nodes of a bredge. Exploiting once more the fact that $\tilde \pi(\tilde g) \simeq \delta(\tilde g - g_*)$ for ER networks with large mean degree, we obtain bredge probabilities as functions of the degrees $k,k'$ of terminal nodes as
\be
b = b(k,k') = 1-\big[1-(1-pg_*)^{k-1}\big]\big[1-(1-pg_*)^{k'-1}\big]
\label{bkkprime}
\ee
 Noting that for a Poisson degree distribution $\frac{k}{c} p_k = p_{k-1}$ we relabel $k\leftarrow k-1$ and $k'\leftarrow k'-1$ on the r.h.s. of Eq.\,\eqref{bkkprime}. Following the reasoning for APs above, we once more exploit the fact that scaled (relabeled) degrees $x = k/c$ and $y=k'/c$ are normally distributed random variables of mean 1 and variance $1/c$. Rewriting Eq.\,\eqref{bkkprime} in terms of scaled relabeled degrees gives
\be
b = b(x,y) = 1-\big[1-(1-pg_*)^{cx}\big] \big[1-(1-pg_*)^{cy}\big]\ .
\ee
We can solve this, for instance, for $y$ to obtain
\be
y = y(b|x) = \frac{1}{c \ln(1-pg_*)} \ln\Bigg[1- \frac{1-b}{1-(1-pg_*)^{cx}}\Bigg]\ ,
\label{yofbgx}
\ee
provided that $b > (1-pg_*)^{cx}$, giving $y$ for any given value of $x$ as a function of the bredge probability $b$. This allows us to transform the Gaussian pdf  $\pi(y) = \sqrt{\frac{c}{2\pi}}\,\exp\big[- \frac{c}{2} (y-1)^2\big]$ of the scaled degree $y$ --- using an analogous transformation of variables identity for pdfs, albeit now conditioned on $x$ --- into
\bea
\pi(b|x) &=& \cN(x)\, \pi(y)\Big|\frac{\partial y}{\partial b}\Big|\, \Theta\Big(b-(1-pg_*)^{cx}\Big)\nn\\
          &=&  \frac{\cN(x)\,\exp\Big[-\frac{c}{2}\Big(y(b|x)-1\Big)^2\Big]}{\sqrt{2\pi c}\,|\ln(1-p g_*)| \, \big[b - (1-pg_*)^{cx}\big] }\nn\\
& & \times \Theta\Big(b-(1-pg_*)^{cx}\Big)\ ,
\eea
with $y(b|x)$ given by Eq.\,\eqref{yofbgx} and $\Theta(x)$ the Heaviside step-function. Here $\cN(x)$ is a normalization factor that is needed due to the $x$ dependent restriction on the allowed range of $b$ values, to ensure that the $\pi(b|x)$ are normalized pdfs for all $x$. It cannot be evaluated in closed form and has to be obtained numerically. This finally gives
\be
\pi(b) = \int \rd x \, \pi(b|x)\,\pi(x)
\label{intpib}
\ee
with $\pi(x)$ the pdf of an $\cN(1,1/c)$ normal random variable given by Eq.\,\eqref{eq:piofx}. 
%(the restricition of the support to $x\ge 1/c$ results in a negligible modification for $c \gg 1$ ). 
The integral in Eq.\,\eqref{intpib}, too, has to be done numerically.

In Sec.\ref{sec:Res} below we show that even for moderate values of the mean degree $c$ these results already provide a decent approximation for the distributions of articulation point probabilities and bredge probabilities.

\begin{figure*}[t!]
\setlength{\unitlength}{1mm}
\begin{picture}(170,60)
\put(-3,5){\includegraphics[width=0.5\textwidth]{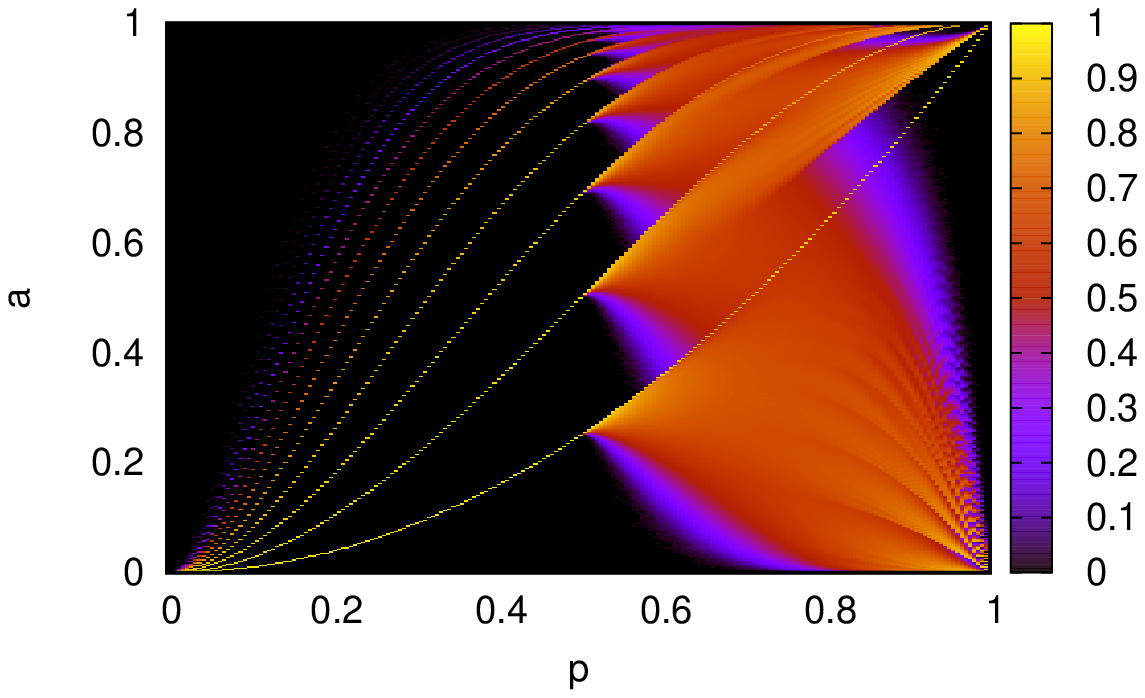}}
\put(87,5){\includegraphics[width=0.5\textwidth]{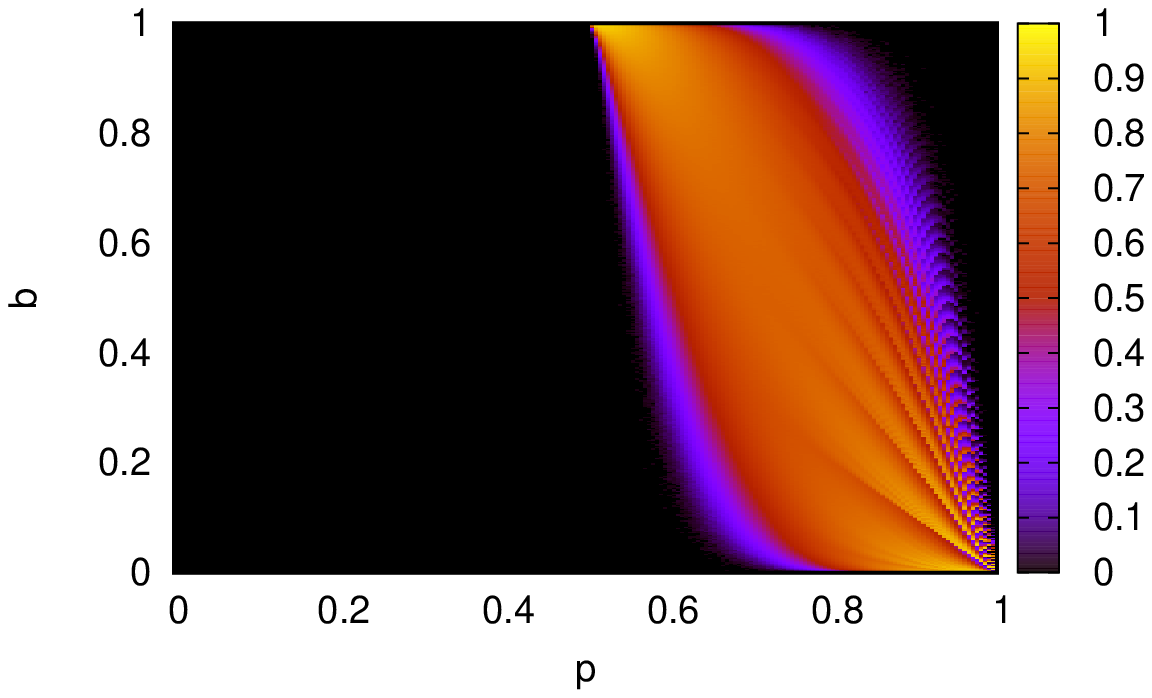}}
\put(-3,58){\large{\bf (a)}}
\put(87,58){\large{\bf (b)}}
\end{picture}
\vspace{-12mm}
\caption{(Color online) {\bf (a)}: Heat-map of the pdf $\pi(a)$ of articulation point probabilities
% and average  articulation point probability (full line)
 for an ER network of mean degree $c=2$. {\bf (b)}: heat-map of the pdf $\pi(b)$ of bredge probabilities 
%and $p$ dependent average bredge- probability 
for the same system. To achieve a color-code that remains discriminate also at relatively low values of pdfs, a nonlinear mapping of pdfs into the interval [0,1] of the form $\pi(\cdot) \to \sqrt{\pi(\cdot)}/(0.4+\sqrt{\pi(\cdot)}\,)$ is adopted.}
\label{fig:ER2piab}
\end{figure*}

\begin{figure*}[ht!]
\setlength{\unitlength}{1mm}
\begin{picture}(170,60)
\put(-3,5){\includegraphics[width=0.5\textwidth]{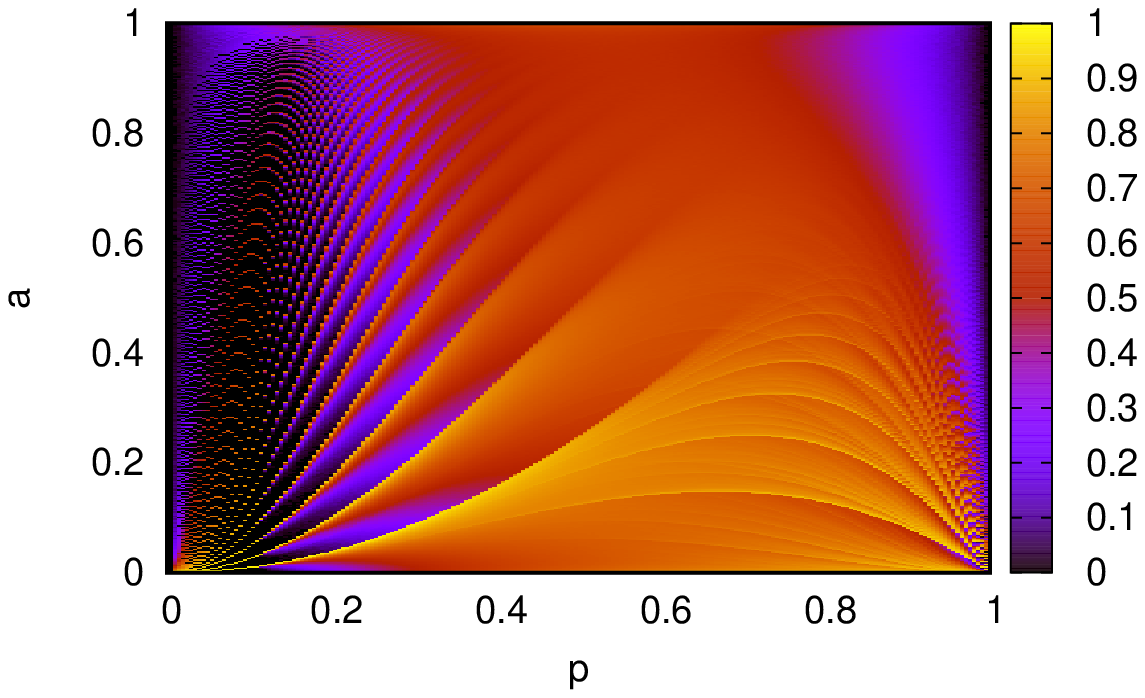}}
\put(87,5){\includegraphics[width=0.5\textwidth]{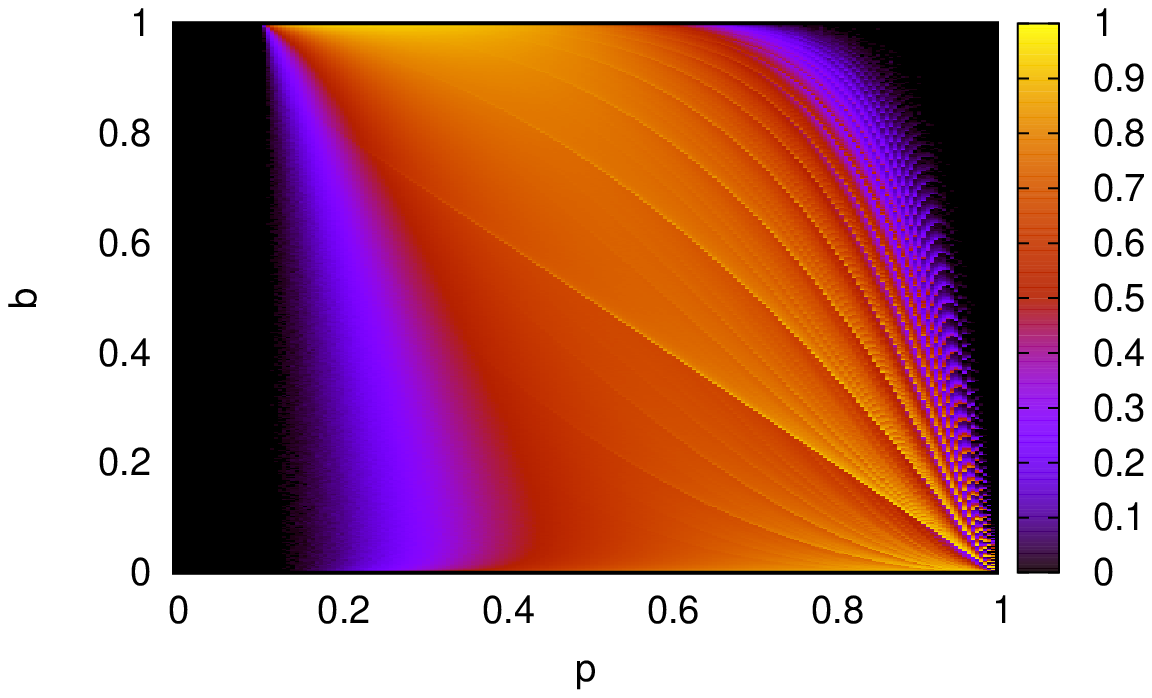}}
\put(-3,58){\large{\bf (a)}}
\put(87,58){\large{\bf (b)}}
\end{picture}
\vspace{-12mm}
\caption{(Color online) {\bf (a)}: Heat-map of the pdf $\pi(a)$ of articulation point probabilities for a scale-free network with degree distribution $p_k \propto k^{-3}$, with $k\ge 2$.  {\bf (b)}: heat-map of the pdf $\pi(b)$ of bredge probabilities for the same system. As in Fig.\,\ref{fig:ER2piab}, the color code is generated using the non-linear mapping $\pi(\cdot) \to \sqrt{\pi(\cdot)}/(0.4+\sqrt{\pi(\cdot)}\,)$.}
\label{fig:pw3m2piab}
\end{figure*}

%%%%%%%%%%%%%%%%%%%%%%%%%%%%%%%%%%%%%%%%%%%%%%%%%%%%%%%%%%
\section{Results}\label{sec:Res}
%%%%%%%%%%%%%%%%%%%%%%%%%%%%%%%%%%%%%%%%%%%%%%%%%%%%%%%%%%
Below we present results both for synthetic networks in the configuration model class in the thermodynamic limit, and for single instances of a real world network.

%%%%%%%%%%%%%%%%%%%%%%%%%%%%%%%%%%%%%%%%%%%%%%%%%%%%%%%%%%
\subsection{Synthetic Networks}
%%%%%%%%%%%%%%%%%%%%%%%%%%%%%%%%%%%%%%%%%%%%%%%%%%%%%%%%%%
Figures \ref{fig:ER2piab} and \ref{fig:pw3m2piab} show heat-maps of the $p$-dependent probability density functions $\pi(a)$ of articulation point probabilities and and $\pi(b)$ of bredge probabilities, respectively. Results in Fig.\, \ref{fig:ER2piab} are for ER networks of mean degree $c=2$, those in Fig.\,\ref{fig:pw3m2piab} for scale free networks with degree distribution $p_k \propto k^{-3}$ for $k\ge 2$. The first prominent difference between the two networks is a different percolation threshold, namely $p_c=0.5$ for the ER network and $p_c \simeq 0.1$ for the scale-free network. Apart from the fact that the scale free network has a lower percolation threshold than the ER network, another qualitative difference is in the fact that it has a minimum degree $k_{\rm min}=2$ and thus it has, at $p=1$, no finite isolated clusters or even sets of nodes that form tree-like structures. This feature explains the main qualitative difference of the heat-maps of AP probabilities in the ER and the scale-free network: non-leaf nodes on trees are {\em always\/} APs; these are abundant in the ER network, explaining the fact that, for large $p$, the pdf $\pi(a)$ of AP probabilities has significant mass for $a \simeq 1$ in the ER case, whereas for $p$ close to 1, the pdf of AP probabilities is very close to zero near $a=1$ in the scale free network where trees do not exist right at $p=1$.

There are, in both figures, sharp $p$-dependent structures that are clearly visible sufficiently far above the percolation threshold. They can be rationalized, both for the heat maps of $\pi(a)$ and $\pi(b)$ in terms of local neighborhoods of nodes and edges, as explained in Sec. \ref{sec:APs} for articulation points, and in Sec \ref{sec:Bs} for bredges respectively.

For articulation points there is a family of curves connecting the points $(p,a)=(0,0)$ and $(p,a)=(1,1)$ described by Eq.\,\eqref{AFCkp} for finite clusters and  by Eq.\,\eqref{AGCkp} with $k_t>0$ for root nodes of trees attached to the GCC. These are clearly visible in the entire $p$ range for the ER network with mean degree $c=2$ as finite clusters and trees attached to the GCC exist at all $p$. For the scale-free network with $k_{\rm min}=2$ there are no finite clusters and tree-like structures at $p=1$. They are, however created by random bond removal, so features with locations described by these equations become more and more prominent as $p$ is lowered. For articulation points which are {\em not\/} root nodes of trees, Eq.\,\eqref{AGCkp} with  $k_t=0$ defines  a family of curves which connect the points $(p,a)=(0,0)$ and $(p,a)=(1,0)$; these curves describe the location of $p$ dependent peaks in pdfs of AP probabilities which are clearly visible close to $p=1$ as they require that the probability to have cavity percolation probabilities $g_j^{(i)}\simeq 1$ is  sufficiently large. This family of curves will thus become blurred and disappear as the percolation transition is approached from above.

For bredges, Eq.\,\eqref{BGCkkpp} defines a family of curves which correspond to maxima in pdfs of bredge probabilities, which --- in analogy to the corresponding family of curves describing AP probabilities ---  will be clearly visible close to $p=1$ as they, too, require the probability to have cavity percolation probabilities $g_j^{(i)}\simeq 1$ to be sufficiently large. As a result, this family of curves will also become blurred and disappear as the percolation transition is approached from above.

\begin{figure*}[t!]
\setlength{\unitlength}{1mm}
\begin{picture}(170,65)
\put(-3,5){\includegraphics[width=0.45\textwidth]{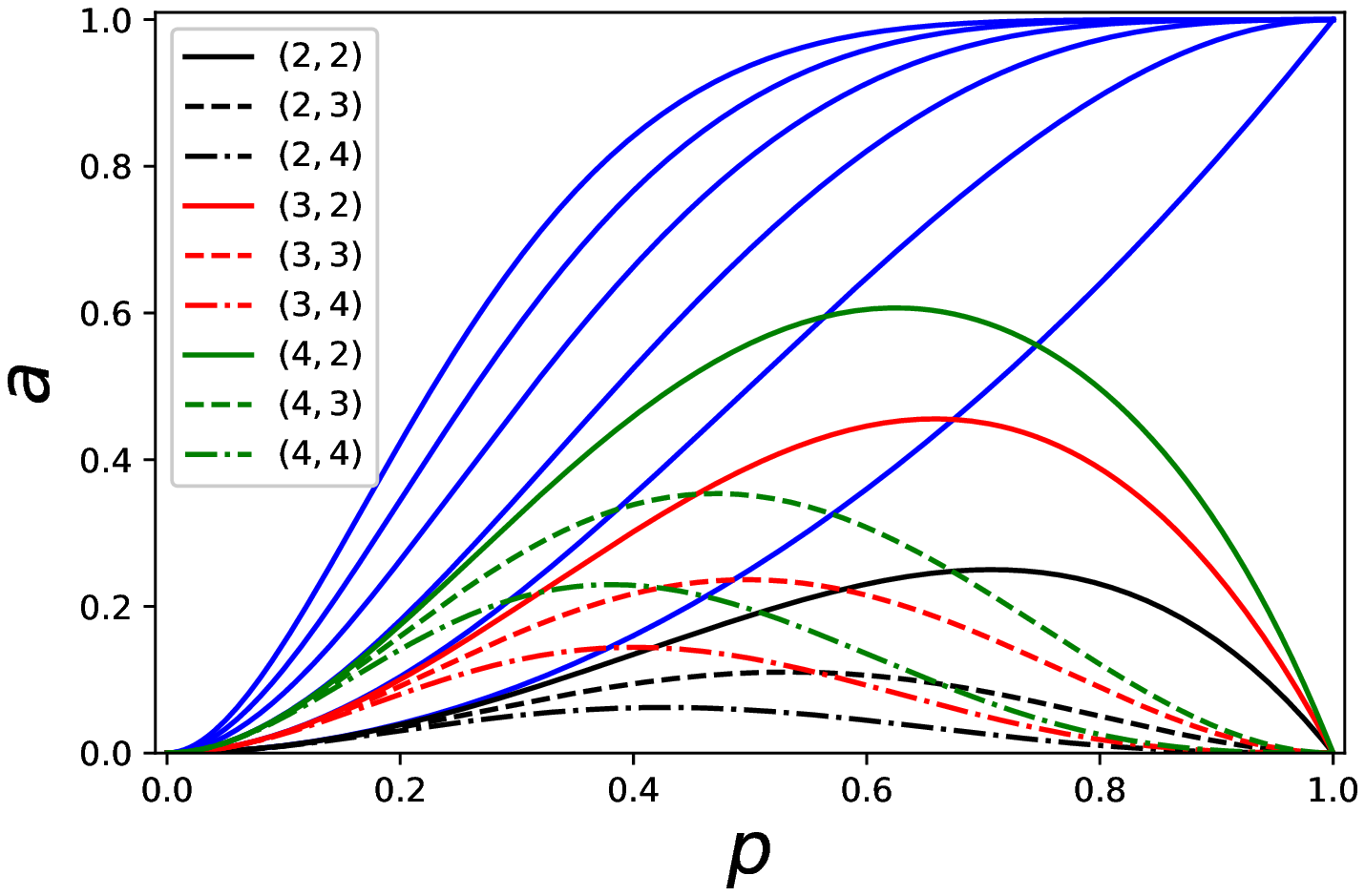}}
\put(87,5){\includegraphics[width=0.45\textwidth]{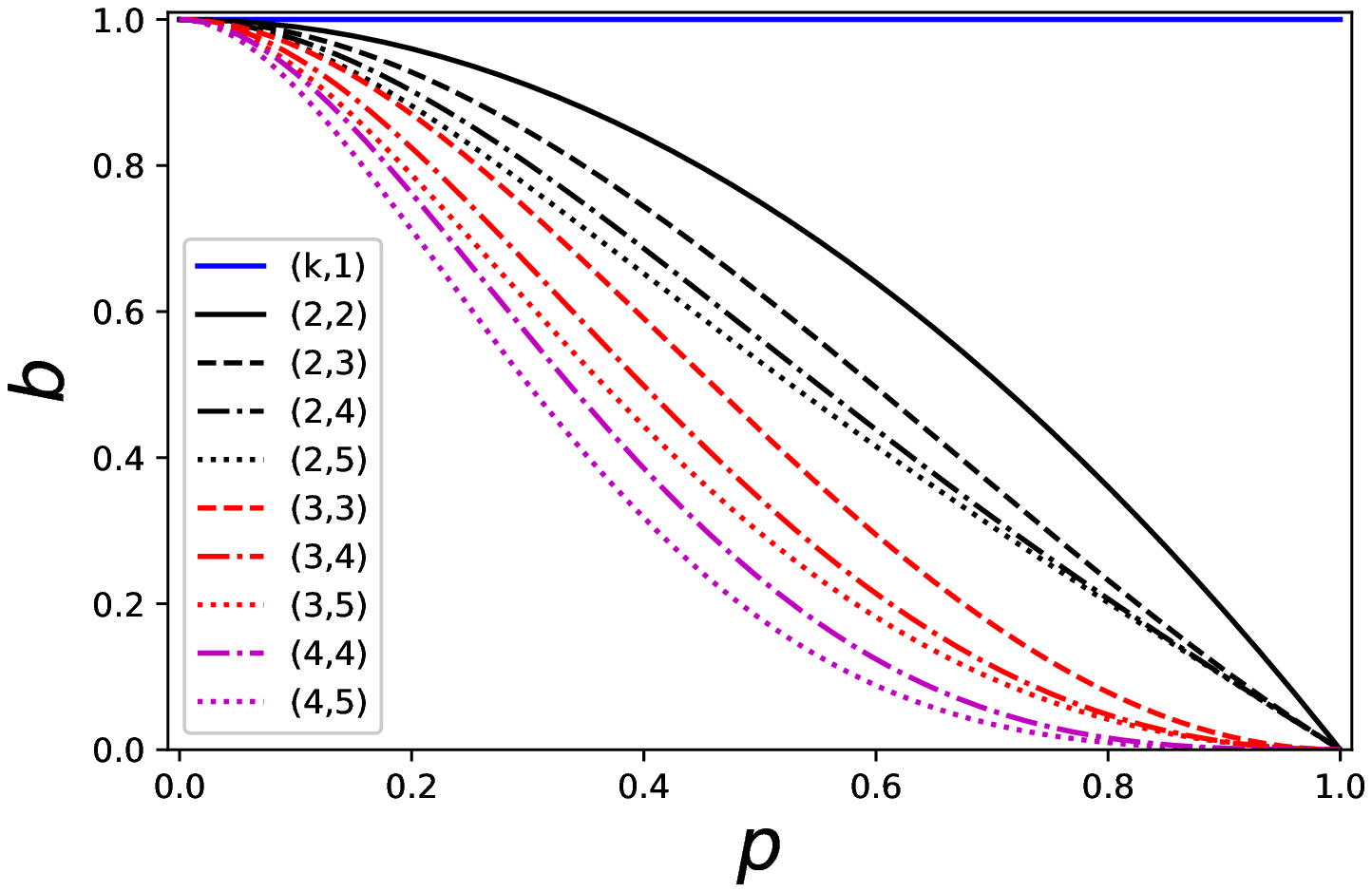}}
\put(-3,55){\large{\bf (a)}}
\put(87,55){\large{\bf (b)}}
\end{picture}
\vspace{-5mm}
\caption{(Color online) {\bf (a)}: Families of $p$-dependent AP probabilities $a_k^{\rm FC}(p)$ (upper unlabeled set of curves, given by Eq.\,\eqref{AFCkp}) for $k=2,\dots,7$ (bottom to top) and $a_{k;k_t,\vk}(p)$ (lower set of curves, given by Eq.\,\eqref{AGCkp});  the lower set of curves corresponds to a sub-family corresponding to a set of nodes of degree $k_i=k$, with $k_t=0$, and $k_j =k'$ for all $j \in \partial i$, and we use the label $(k,k')$  members of this sub-family.  {\bf (b)}: Families of $p$-dependent bredge probabilities of the form $b_{k,k'}(p)$, given by Eq.\,\eqref{BGCkkpp}.}
\label{fig:APB-curves}
\end{figure*}

Some of the curves described by Eqs. \eqref{AFCkp} and \eqref{AGCkp} for APs and  by Eq.\,\eqref{BGCkkpp} for bredges are shown in Fig.\,\ref{fig:APB-curves}, and they agree well with features seen in the heat-maps in Figs.\,\ref{fig:ER2piab} and \ref{fig:pw3m2piab}. Other prominent features seen in the heat maps are {\em not\/} captured by these families of curves, as they require an analysis of local environments of nodes and edges beyond first coordination shells, as discussed and exploited before in \cite{Tishby+18, KuvM20} to rationalize prominent features in pdfs of percolation probabilities.

To the extent that there is some degree of similarity of the results for articulation points and bredges in both cases, it is due to the intimate connection which exists between bredges and articulation points, as  each end-node of a bredge is an articulation point if it is not a leaf node and each articulation point of degree $k\ge 2$ must  have at least one bredge emanating from it \cite{Bonneau+20}. 

\begin{figure*}[ht!]
\setlength{\unitlength}{1mm}
\begin{picture}(170,65)
\put(-3,5){\includegraphics[width=0.45\textwidth]{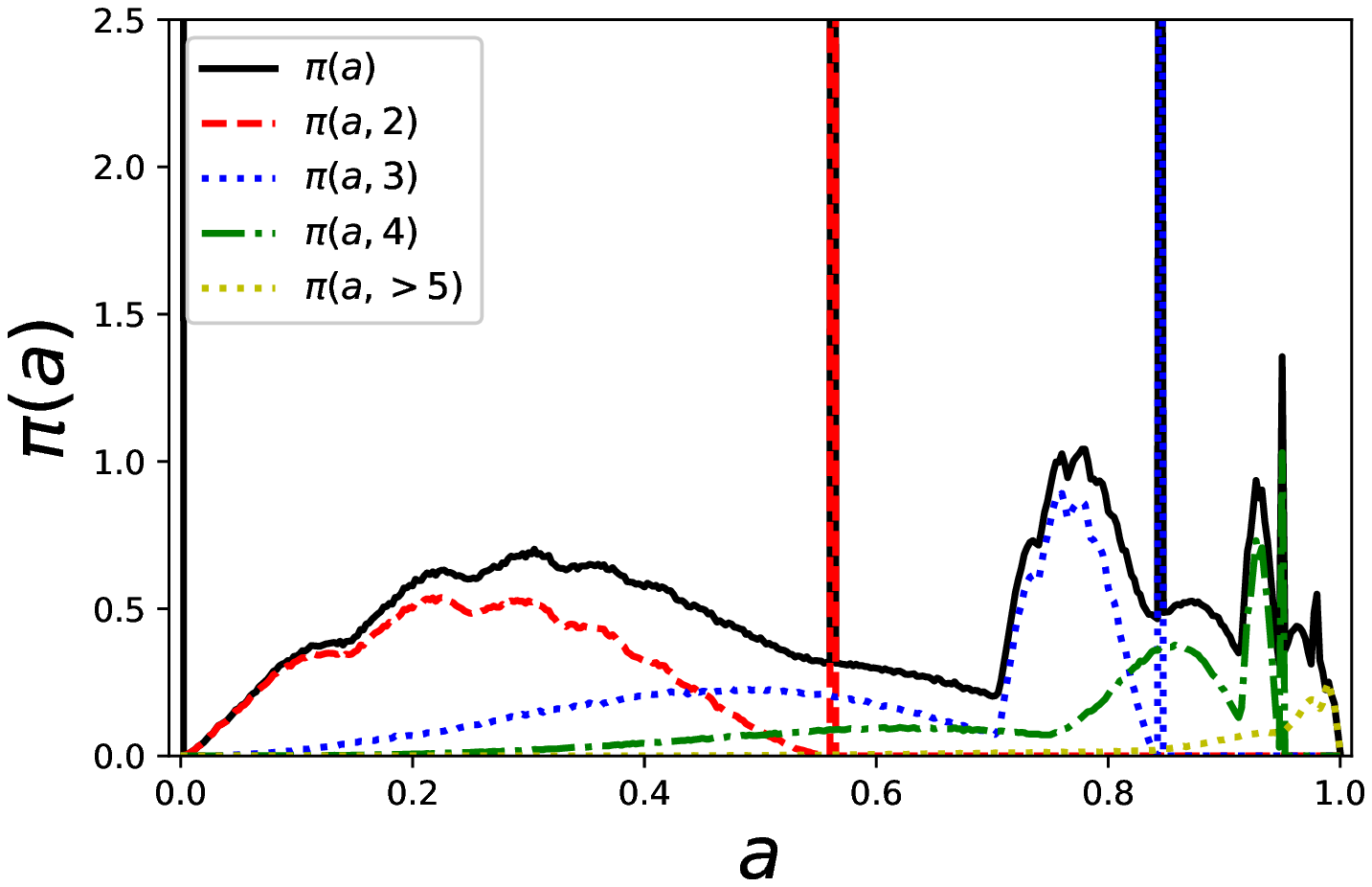}}
\put(87,5){\includegraphics[width=0.45\textwidth]{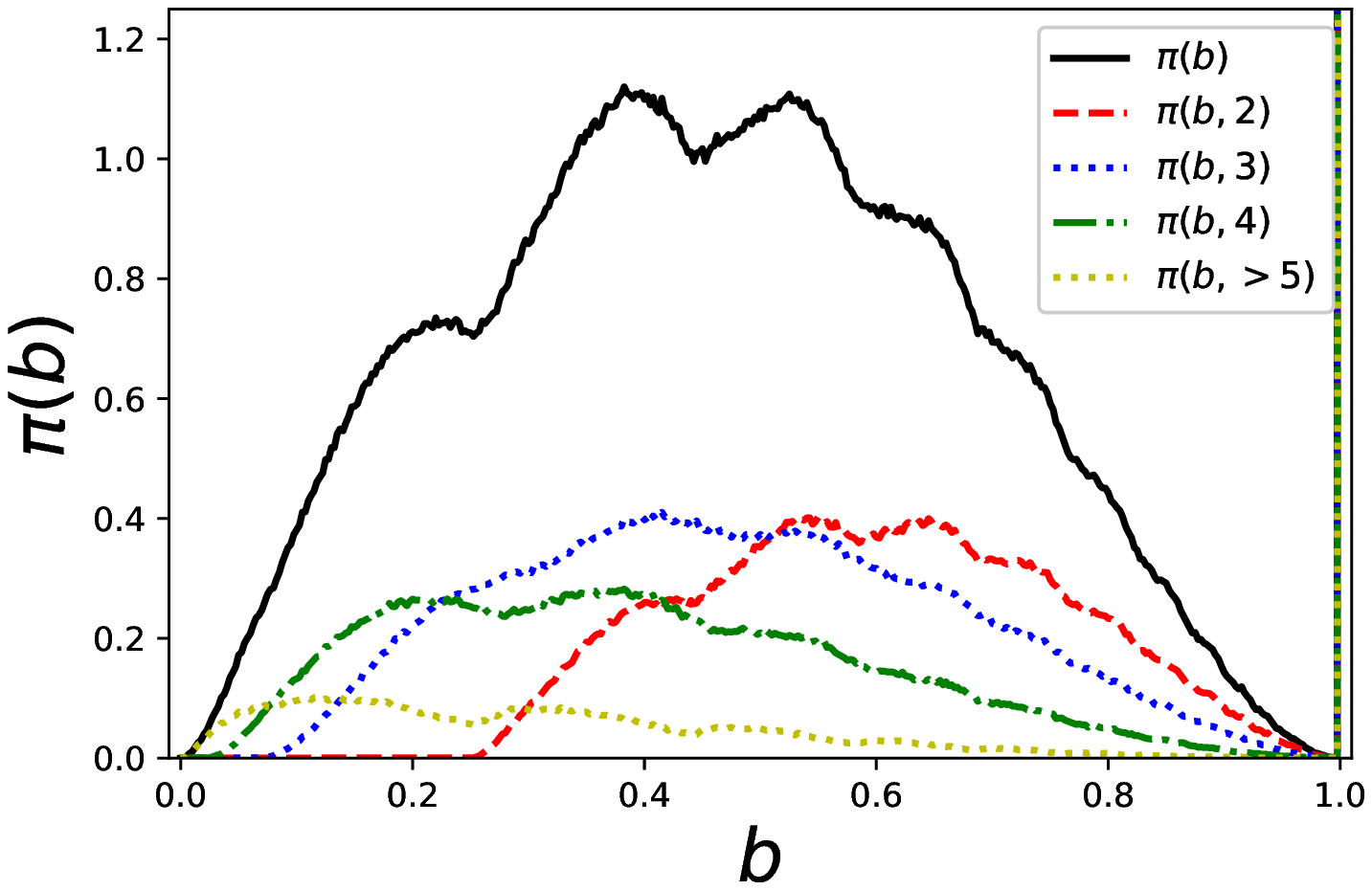}}
\put(-3,55){\large{\bf (a)}}
\put(87,55){\large{\bf (b)}}
\end{picture}
\vspace{-10mm}
\caption{(Color online) {\bf (a)}: Distribution $\pi(a)$ of articulation point probabilities (black full line) and contributions $\pi(a,k)$ for a selection of different degrees, shown here for $k=2, 3,  4$ and $k > 5$,  for the ER network of mean degree $c=2$ at bond retention probability $p=0.75$, with individual contributions decreasing with increasing $k$.  {\bf (b)}: distribution $\pi(b)$ of bredge probabilities (black full line) and contributions $\pi(b,k)=\sum_{k'} \pi(b,k,k')$ for $k= 2, 3, 4$, and $k > 5$ for the same system. Contributions peak at lower $b$ values for increasing $k$. For $\pi(a)$ the $\delta$-peak at $a=0$ originates from nodes with $k=0$ and $k=1$, whereas the $\delta$-peak at $a=1$ is due to nodes on finite clusters and on tree branches that are not leaf nodes. For $\pi(b)$ the $\delta$-peak at $b=1$ is  due to edges which reside on tree branches.}
\label{fig:ER2-cut-dec}
\end{figure*}%%% Fig 4

\begin{figure*}[ht!]
\setlength{\unitlength}{1mm}
\begin{picture}(170,65)
\put(-3,5){\includegraphics[width=0.5\textwidth]{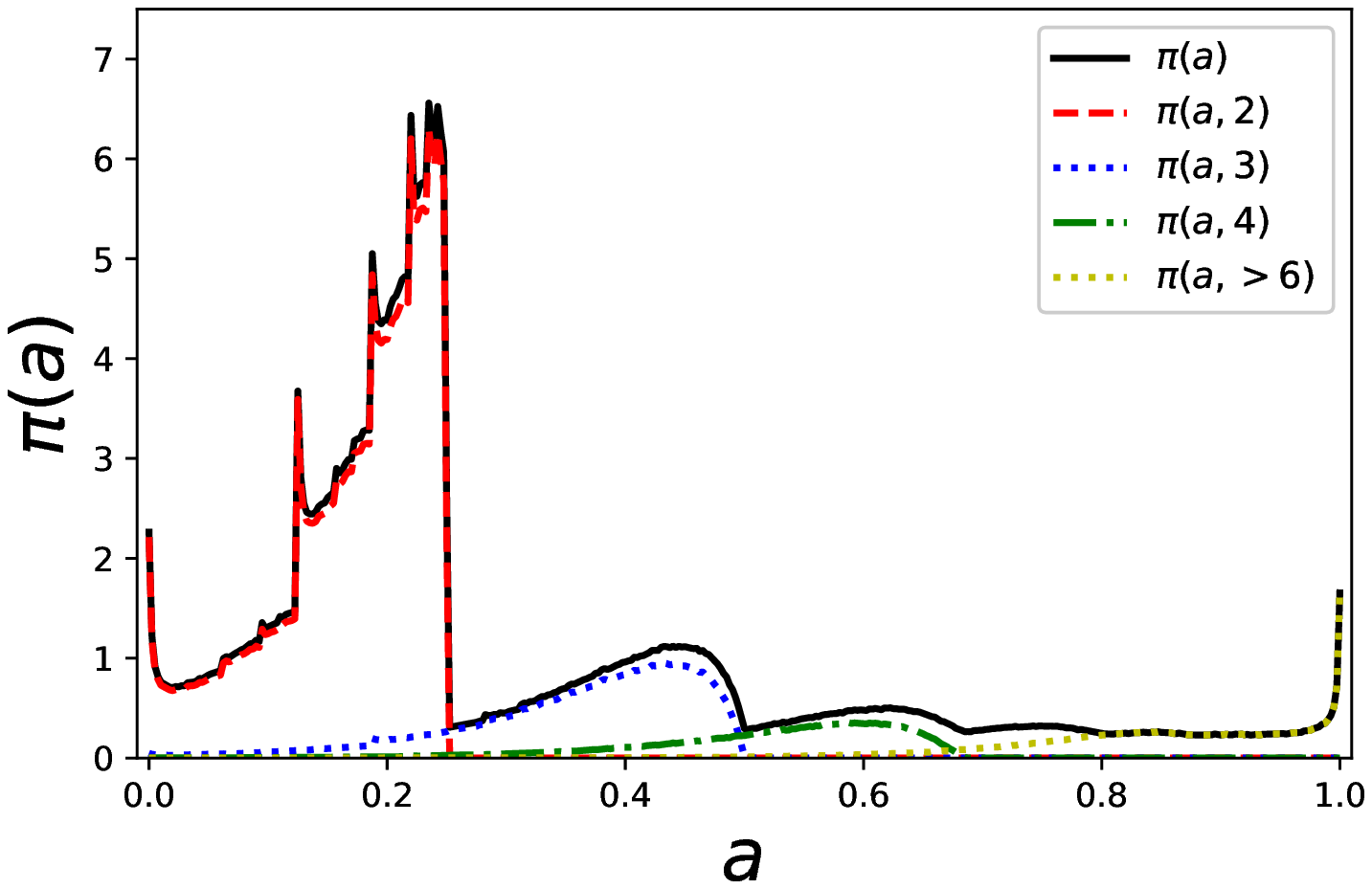}}
\put(87,5){\includegraphics[width=0.5\textwidth]{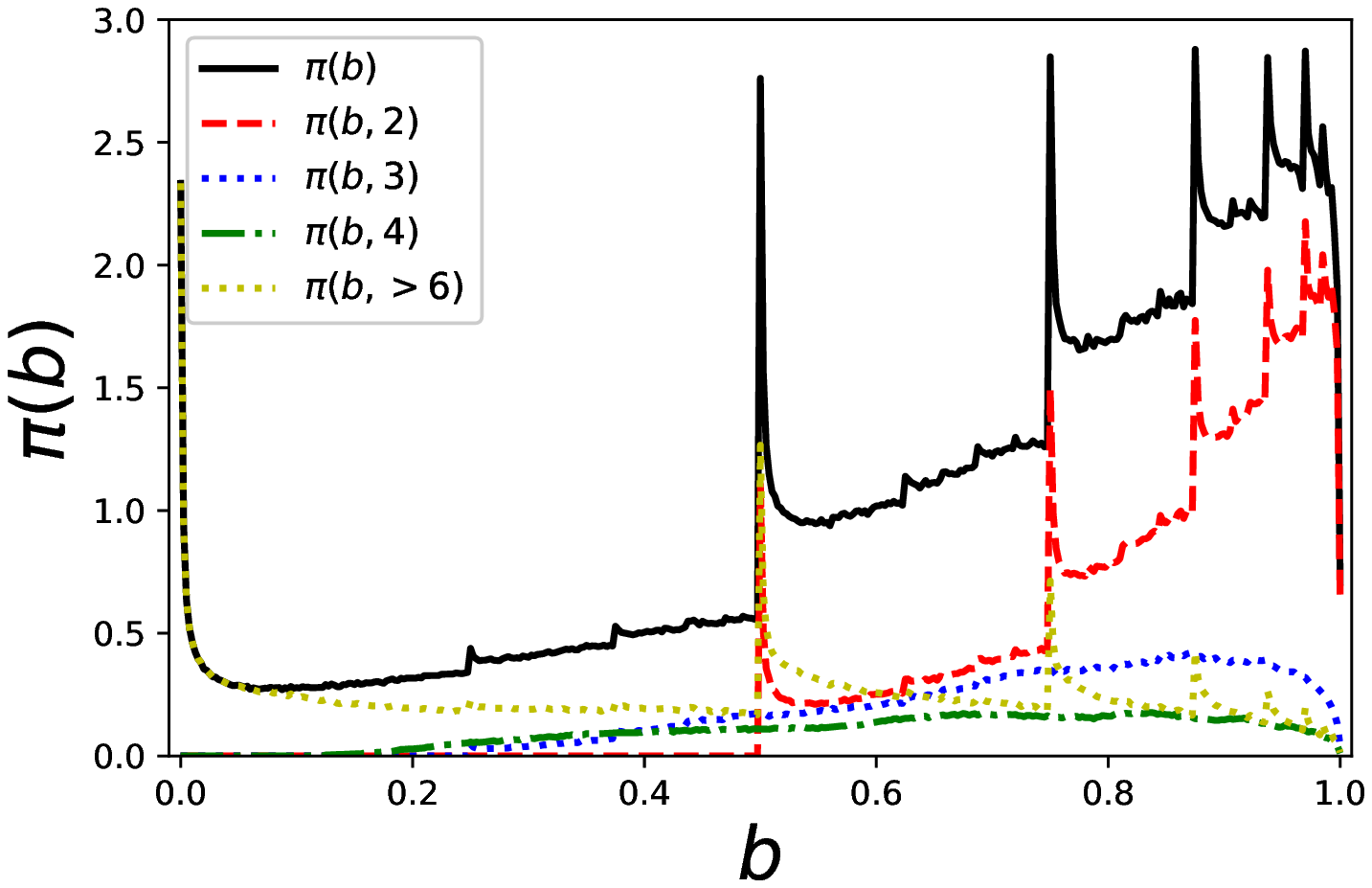}}
\put(-3,60){\large{\bf (a)}}
\put(87,60){\large{\bf (b)}}
\end{picture}
\vspace{-10mm}
\caption{(Color online) {\bf (a)}:  Distribution $\pi(a)$ of articulation point probabilities (black line) and contributions $\pi(a,k)$ for a selection of  different degrees, shown here for $k=2, 3,  4$ and $k > 6$,  for the scale free network at bond retention probability $p=0.5$, with individual contributions decreasing with increasing $k$. Note that typical probabilities of nodes to be APs appear to {\em increase\/} with the degree $k$. {\bf (b)}: distribution $\pi(b)$ of bredge probabilities and contributions $\pi(b,k)=\sum_{k'} \pi(b,k,k')$ for $k=2, 3, 4$  and $k > 6$ for the same system. Contributions peak at lower $b$ values for increasing $k$.}
\label{fig:pw3m2-cut-dec}
\end{figure*}%%% Fig 5

Nonetheless, there are marked differences which are more clearly brought out in plots of $\pi(a)$ and $\pi(b)$ at {\em given\/} values of bond retention probabilities $p$, which are better suited to demonstrate quantitative details. Figure \ref{fig:ER2-cut-dec} shows results for ER networks of Fig.\,\ref{fig:ER2piab} at $p=0.75$. There are $\delta$-peaks at 0 as well as a set of other $\delta$-peaks at locations given by $a_k^{\rm FC}(p=0.75)$  in $\pi(a)$ for $k\ge 2$. The peak at zero is due to  isolated nodes and nodes that are leaf nodes after random bond removal. The continuum part of $\pi(a)$ is due to nodes originally on the 2-core of the GCC. Individual contributions $\pi(a,k)$ for some $k$ are also shown and are seen to contribute to different features in the overall pdf. The pdf $\pi(b)$ of bredge probabilities has a  $\delta$-peak at 1,  originating from finite tree-like clusters or trees attached to the GCC on the original network, whereas the continuous part of $\pi(b)$ is due to nodes on the giant cluster that were part of one or several loops on the GCC of the original network. Any marked features that would occur at locations given by $b_{k,k'}(p=0.75)$ are already blurred. The deconvolution of $\pi(b)$ according to the degree  $k$ of one of the terminal nodes shows that larger values of $k$ entail that bredge probabilities are shifted towards smaller values.

Figure \ref{fig:pw3m2-cut-dec} shows results  for the scale-free network of Fig.\,\ref{fig:pw3m2piab} at $p=0.5$. The scale free network was constructed with $k_{\rm min}=2$, so there are no isolated nodes nor leaves, nor finite tree-like clusters or trees attached to the giant cluster at $p=1$. As a consequence the $\delta$-peaks in $\pi(a)$ and $\pi(b)$ are absent in this system. The deconvolution according to degree reveals that nodes of degree $k=2$ are predominantly responsible for the larger $\pi(a)$ values for $a\le 0.25$ with the peaks at $a=0.1875$ and at $a=0.125$ predicted by Eq.\,\eqref{AGCkp}, while others require to include further coordination shells in the analysis. The sharp maxima in $\pi(b)$ are mainly due to edges, with one terminal node of degree $k=2$. The peaks at $b=0.75$ and at $a=0.5$ are predicted by Eq.\,\eqref{BGCkkpp}, the former as $b(2,2,p=0.5)$ and the latter as an accumulation point of  the $b(k,k',p=0.5)$ for $k'\to \infty$. For the scale-free network, large degrees do occur with sufficient probability to give sufficient weight to this peak. Other sharp peaks at larger values of $b$ can be rationalized by including higher  coordination shells in the analysis.

\begin{figure*}[ht!]
\setlength{\unitlength}{1mm}
\begin{picture}(170,62)
\put(-3,5){\includegraphics[width=0.495\textwidth]{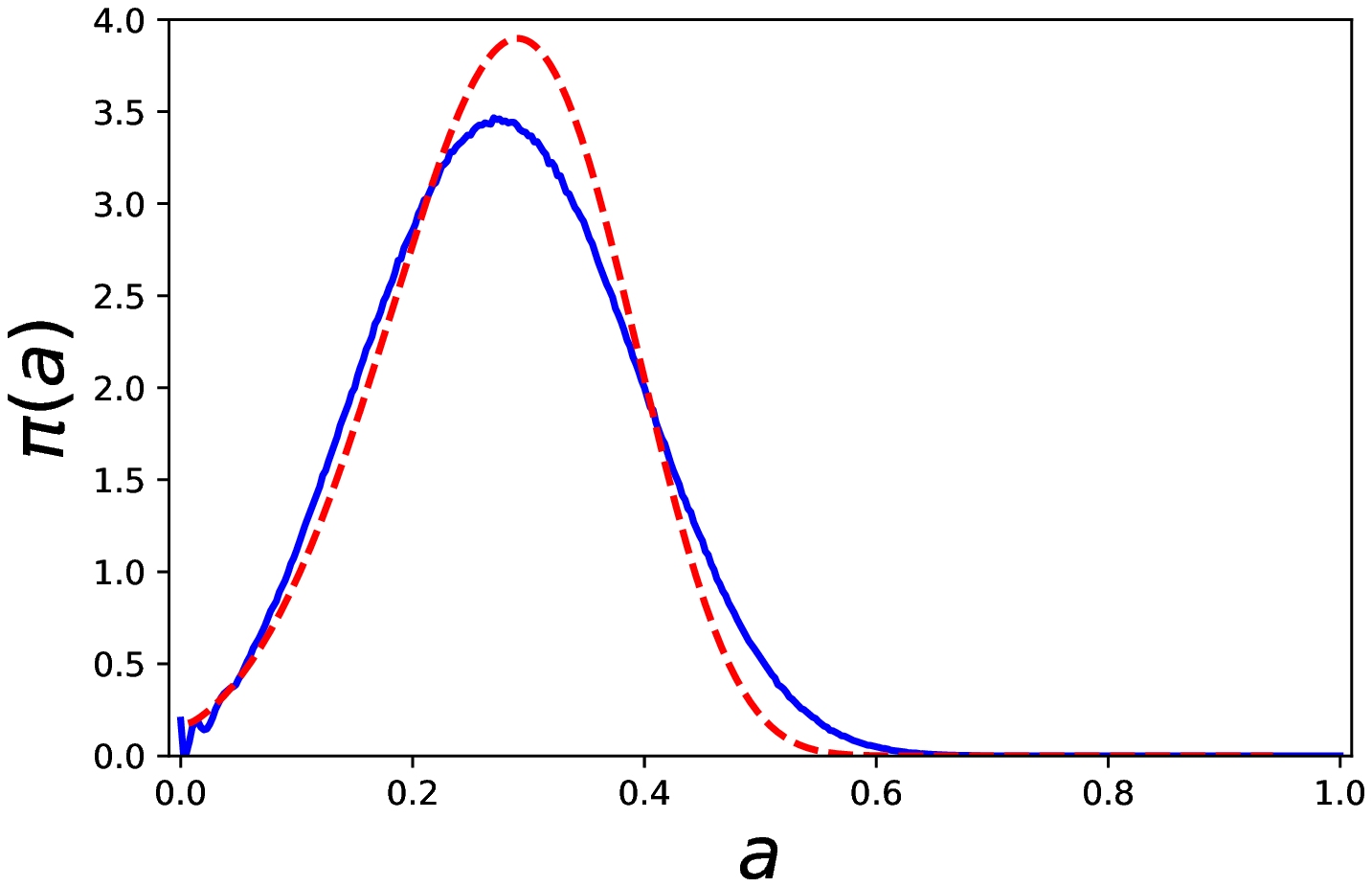}}
\put(87,5){\includegraphics[width=0.5\textwidth]{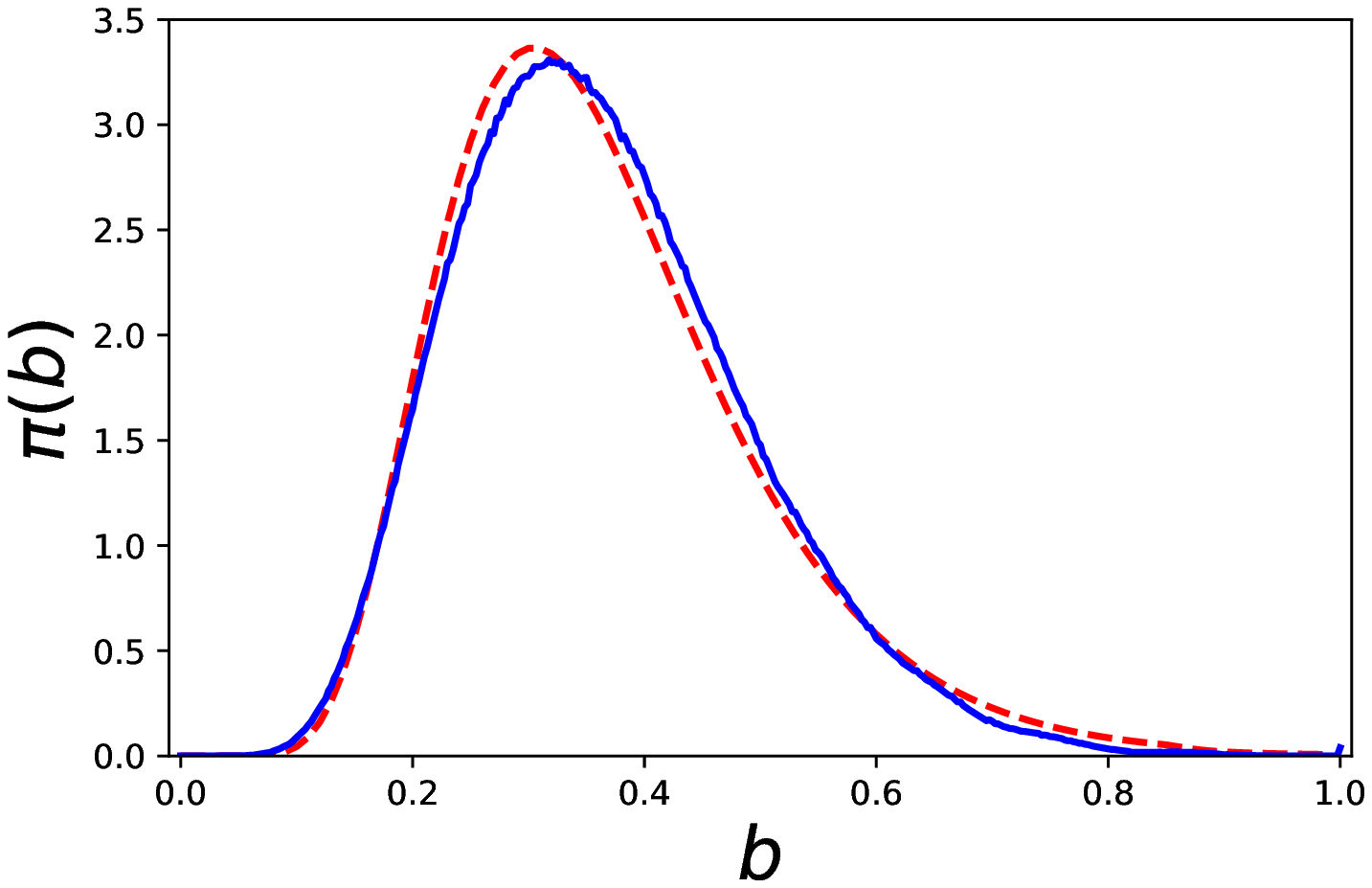}}
\put(-3,60){\large{\bf (a)}}
\put(87,60){\large{\bf (b)}}
\end{picture}
\vspace{-10mm}
\caption{(Color online) {\bf (a)}: Comparing the distribution $\pi(a)$ of articulation point probabilities (blue full line)  with its large mean degree approximation red (dashed line) for an ER graph of mean degree $c=10$ and bond-retention probability $p=0.2$. The analogous comparison for the distribution $\pi(b)$ of bredge probabilities is in panel {\bf (b)}. The agreement between the results of the full cavity analysis and its large mean degree approximation is surprisingly good already  for the moderate value of the mean degree considered here.}
\label{fig:approx-piab}
\end{figure*}

One can harness techniques of  \cite{Ku16} used originally to disentangle  contributions to sparse random matrix spectra coming from the giant connected components and from finite clusters to investigate probabilities of nodes to be articulation points  and probabilities of edges to be bredges {\em conditioned\/} on these nodes and edges having belonged to the GCC prior to any percolation experiment. This is obviously the most relevant issue to study when thinking of maintaining functionality of a network or conversely of attack strategies which would efficiently undermine such functionality. Technically this is done by analysing the message passing equations for two `replica' of indicator variables, one for the system without random bond removals, and one for the same system with random bond removals. This analysis quantitatively confirms the attribution of features in $\pi(a)$ and $\pi(b)$ discussed above according  to whether they are due to nodes or edges originally on the GCC or on finite clusters.

Figure \ref{fig:approx-piab} finally demonstrates that the large mean degree approximation is remarkably efficient already for the fairly moderate value $c=10$ for the mean degree.

\begin{figure*}[ht!]
\setlength{\unitlength}{1mm}
\begin{picture}(170,62)
\put(-3,5){\includegraphics[width=0.5\textwidth]{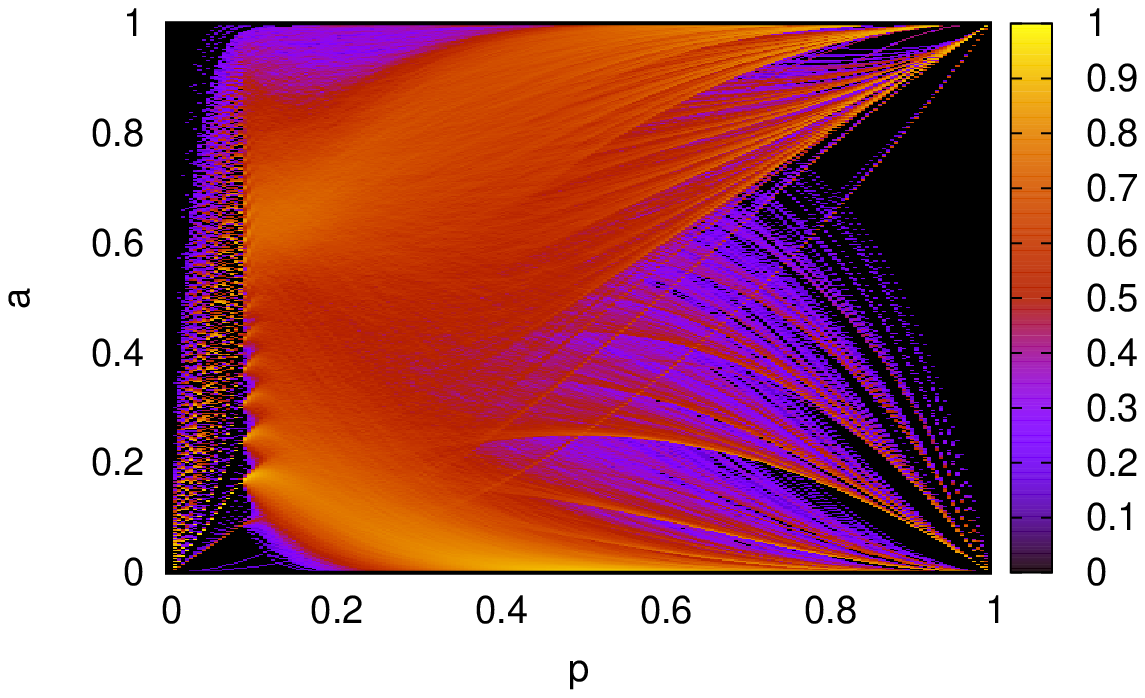}}
\put(87,5){\includegraphics[width=0.5\textwidth]{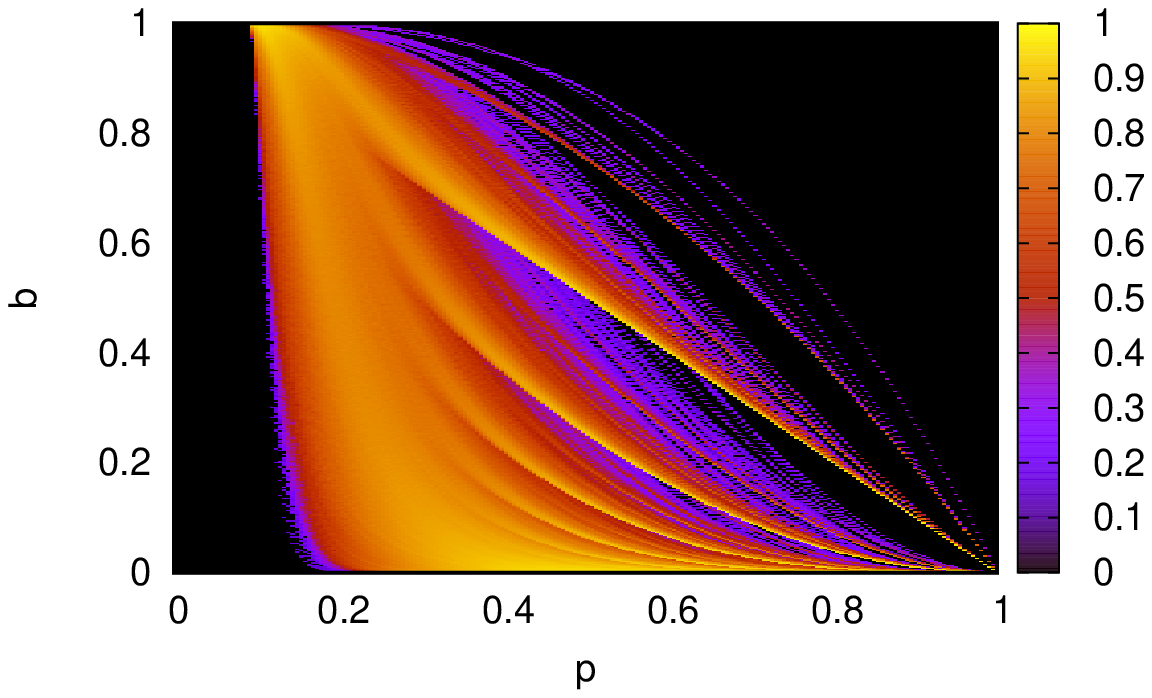}}
\put(-3,58){\large{\bf (a)}}
\put(87,58){\large{\bf (b)}}
\end{picture}
\vspace{-10mm}
\caption{(Color online) {\bf (a)}: Heat-map of the pdf $\pi(a)$ of articulation point probabilities for the Gnutella file-sharing network. {\bf (b)}: heat-map of the pdf $\pi(b)$ of bredge probabilities for the same system. The non-linear map to produce colour-codes is the same as in Figs.\,\ref{fig:ER2piab} and \ref{fig:pw3m2piab}.}
\label{fig:gnutella-piab}
\end{figure*}

%%%%%%%%%%%%%%%%%%%%%%%%%%%%%%%%%%%%%%%%%%%%%%%%%%%%%%%%%%
\subsection{Single-Instance Cavity for a Real World Network}
%%%%%%%%%%%%%%%%%%%%%%%%%%%%%%%%%%%%%%%%%%%%%%%%%%%%%%%%%%
If we compare results for synthetic networks shown above with those for a real world network, we can state the following. There are $p$-dependent structures --- with the {\em same\/} $p$-dependence --- in the real world and synthetic networks. In Fig.\,\ref{fig:gnutella-piab} we present results for one such real world network, a symmetrized version of the Gnutella file-sharing network\cite{Lesk14} with $N= 62,586$ nodes. It is notable that the clarity with which individual features show up depends very much on the system, and differences between the networks are clearly visible. For example, except in the vicinity of the percolation transition at $p_c \simeq 0.1$, the pdf of bredge probabilities is concentrated at significantly lower values of $b$ in the Gnutella network than in the two examples of synthetic random networks. The same appears to be true for the pdf of articulation point probabilities generated by nodes that are on the 2-core of the original network, for which therefore $a_i(p) \to 0$ as $p\to 1$.

\begin{figure*}[t!]
\setlength{\unitlength}{1mm}
\begin{picture}(170,118)
\put(-3,60){\includegraphics[width=0.46\textwidth]{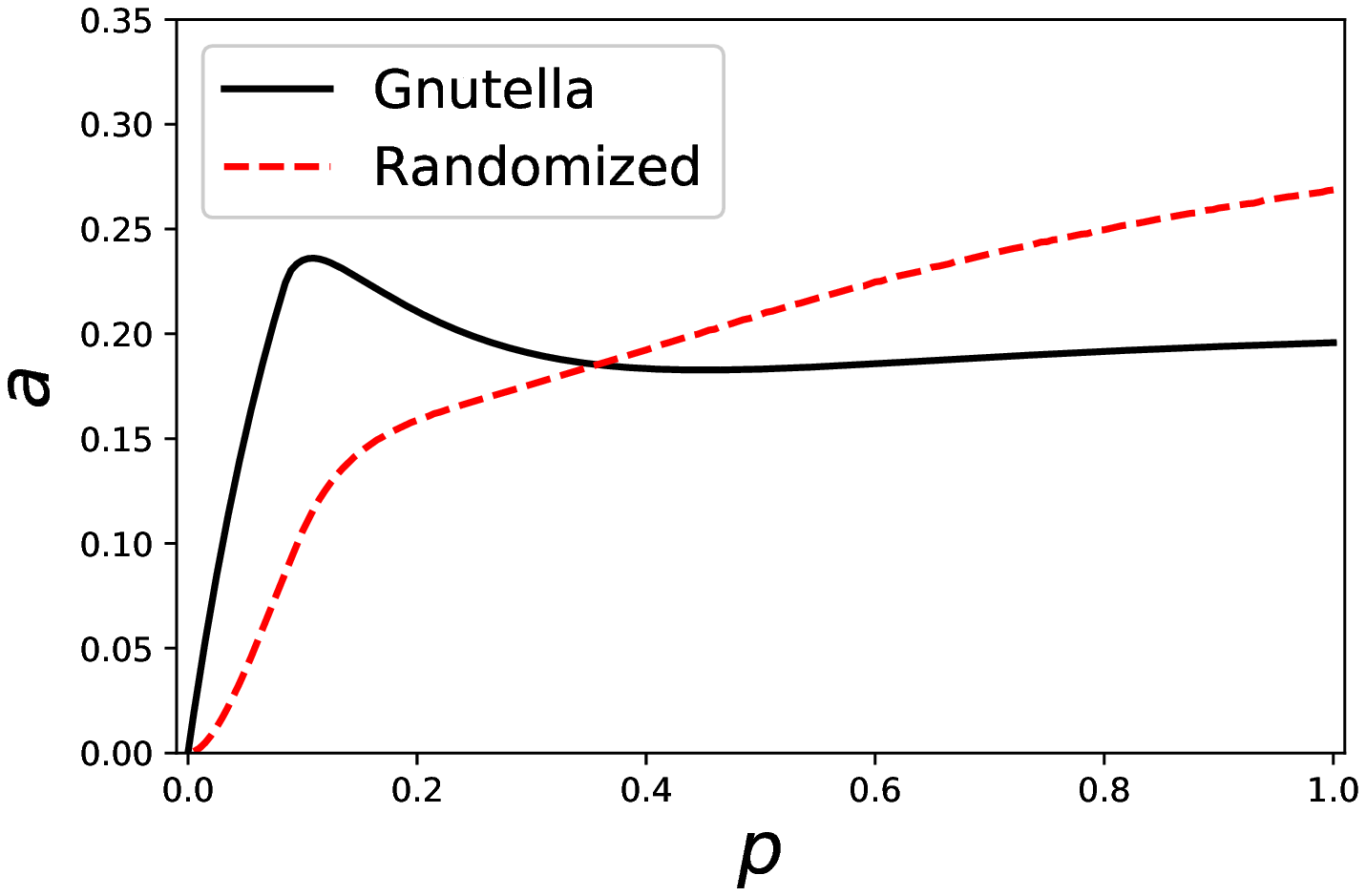}}
\put(87,60){\includegraphics[width=0.46\textwidth]{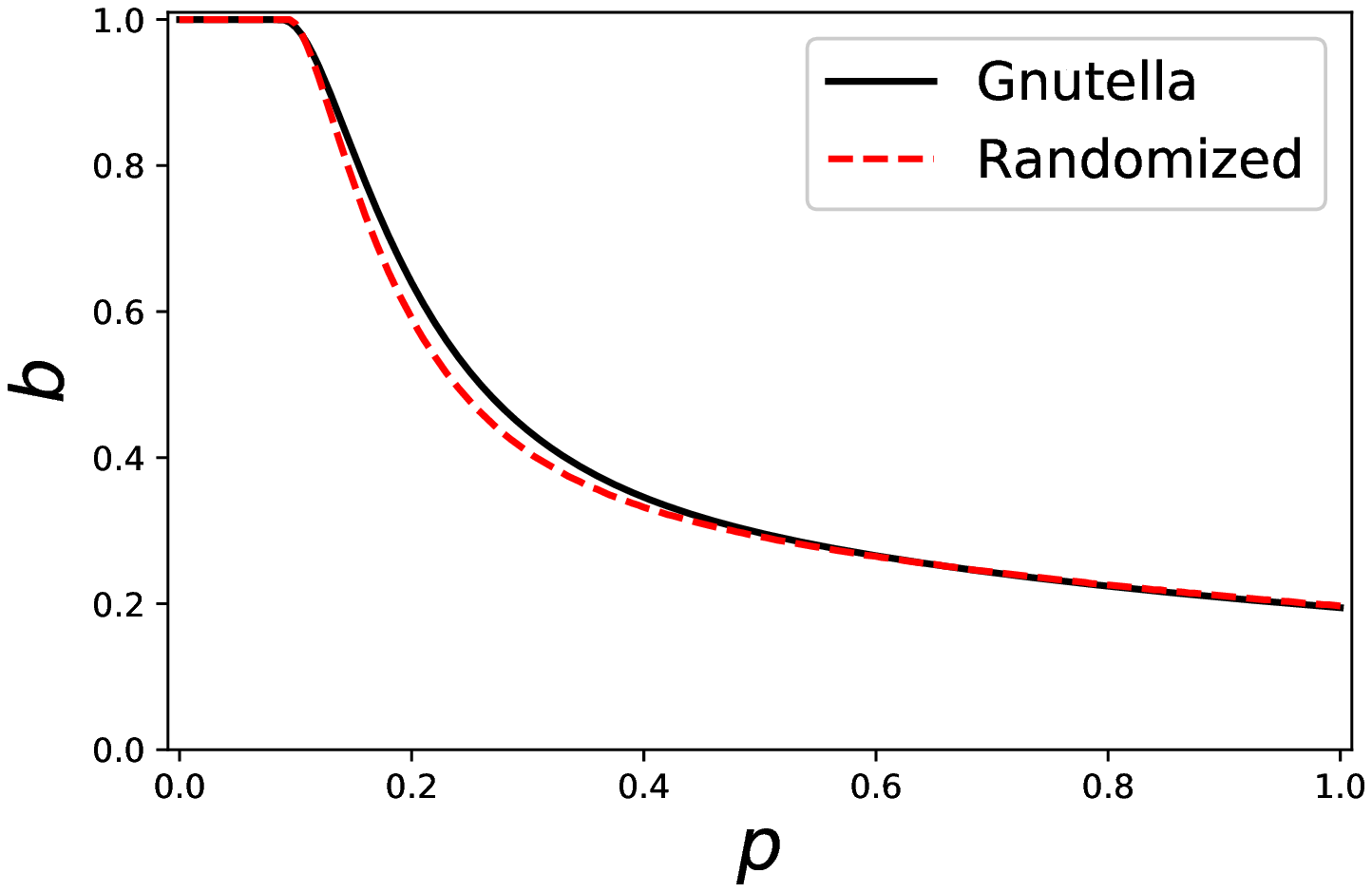}}
\put(-3,7){\includegraphics[width=0.46\textwidth]{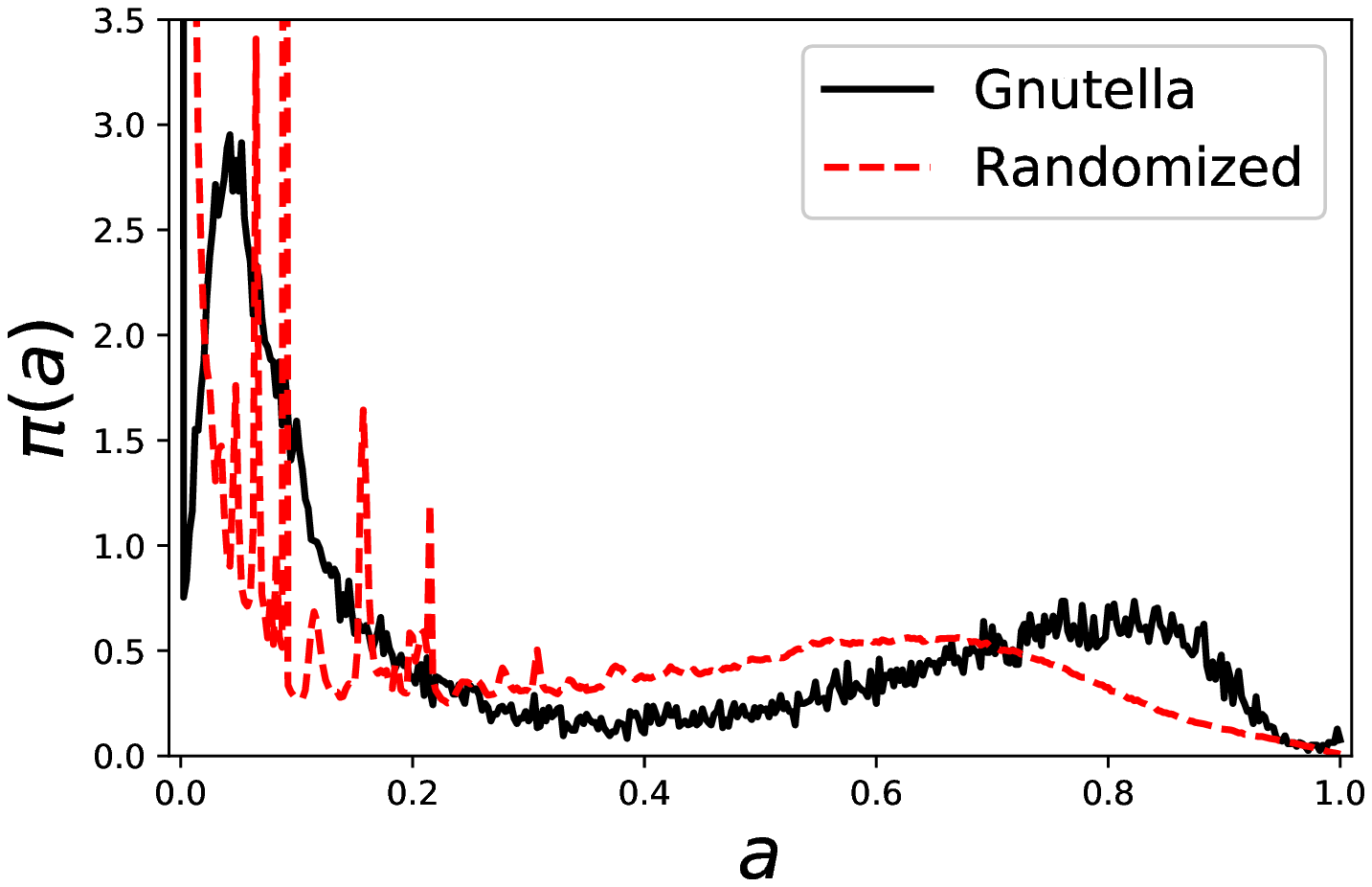}}
\put(87,7){\includegraphics[width=0.46\textwidth]{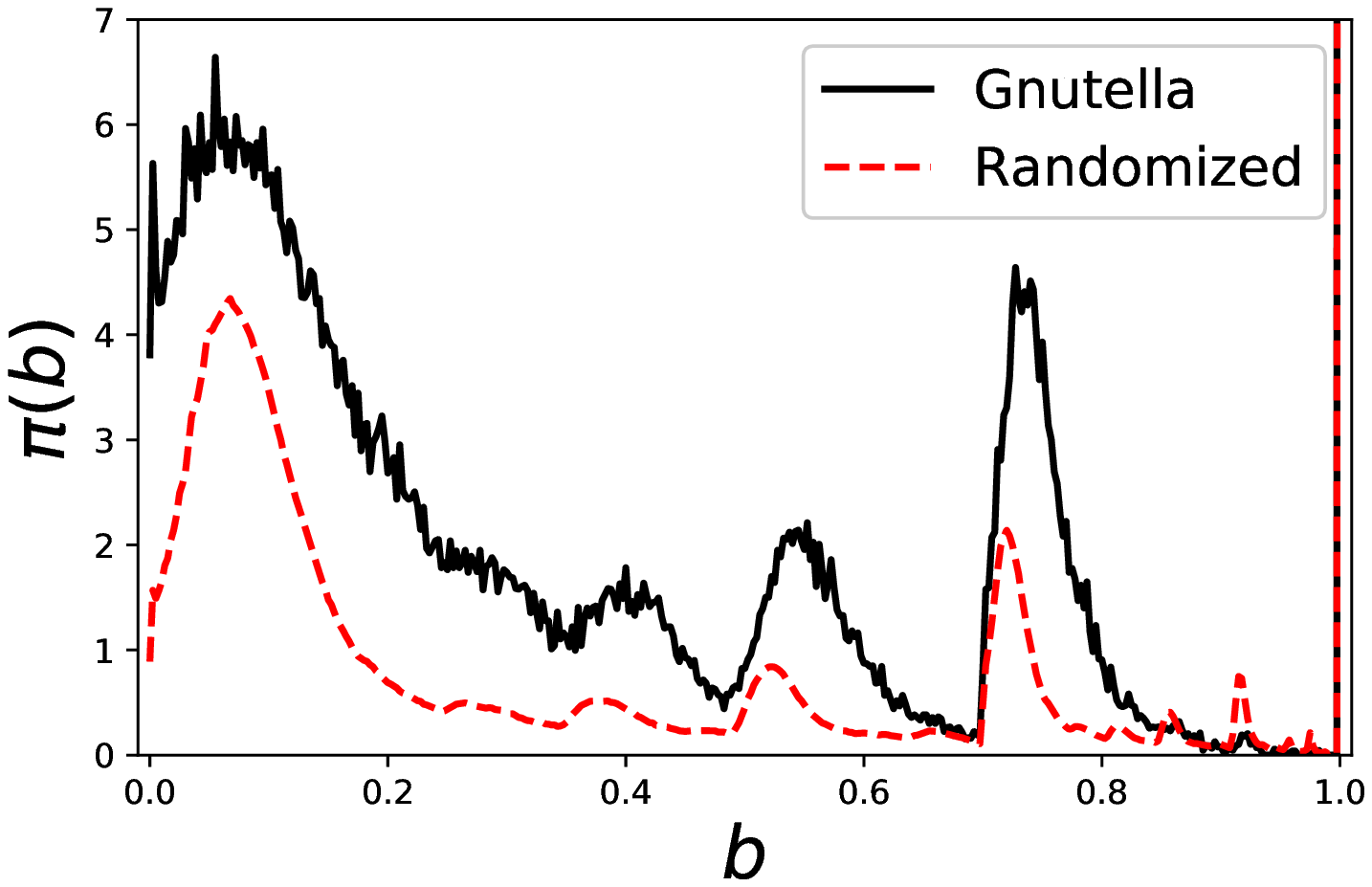}}
\put(-3,110){\large{\bf (a)}}
\put(87,110){\large{\bf (b)}}
\put(-3,57){\large{\bf (c)}}
\put(87,57){\large{\bf (d)}}
\end{picture}
\vspace{-10mm}
\caption{(Color online) {\bf (a)} Average AP probabilities for the Gnutella file sharing network undergoing a percolation process as a function of bond retention probability $p$, compared with the $p$-dependent AP probabilities for a randomized version of the same network. {\bf (b)} The same comparison for the $p$-dependent average bredge probabilities. Panels {\bf (c)} and {\bf (d)} compare full distributions $\pi(a)$ of AP probabilities and $\pi(b)$ of bredge probabilities for the original and the randomized version of the Gnutella network at bond retention probability $p=0.3$. While at the level of average AP and bredge probabilities the Gnutella network and its randomized version are very similar, the full pdfs $\pi(a)$ and $\pi(b)$ exhibit remarkable differences.}
\label{fig:GnutellaAndToConfig}
\end{figure*}

A recent study of bredges in real world networks \cite{Wu+18} has revealed that the fraction of bredges in these networks was very close to the fraction in randomized networks with the same degree distribution. In Fig.\,\ref{fig:GnutellaAndToConfig} we confirm this for the Gnutella network undergoing random bond removal for the {\em average\/} bredge probabilities at all values of the bond retention probability $p$, and to a lesser degree of similarity for average AP probabilities. However, despite the closeness of values of {\em average\/} AP and bredge probabilities in the original network and its randomized version, there remain marked differences at the level of the full distributions $\pi(a)$ and $\pi(b)$ of AP and bredge probabilities, respectively.

%%%%%%%%%%%%%%%%%%%%%%%%%%%%%%%%%%%%%%%%%%%%%%%%%%%%%%%%%%
\subsection{Single-Instance Cavity Algorithm to Identify Articulation Points and Bredges}
%%%%%%%%%%%%%%%%%%%%%%%%%%%%%%%%%%%%%%%%%%%%%%%%%%%%%%%%%%
The message passing approach can also be used to identify individual APs and bredges in a given network, rather than local AP probabilities and bredge probabilities in ensembles of systems affected by random bond or site removal. To this end, one returns to the message passing equations  for the indicator variables $n_i$ and the cavity indicator variables $n_j^{(i)}$ for a single realization of the bond-occupancy variables $x_{ij}$, i.e.,  Eqs.\,\eqref{ni-eq}  and  \eqref{nji-eq}. An iterative solution of Eqs.\,\eqref{nji-eq} for $x_{ij}\equiv 1$ would allow one to identify  APs and bredges in the original network via Eq.\,\eqref{hatni} and  Eq.\,\eqref{nij}, respectively. Given any realization of the $x_{ij}$ resulting from random or targeted bond removal one could repeat that analysis and find bredges and APs in the same manner {\em after\/} bond removal. 

We have tested this idea for the Gnutella file-sharing network. Although the message passing  algorithm will be exact only on trees, we found it to be extremely efficient and accurate for this real world network, despite the fact that the network contains a large number of loops. For bredges, comparison with exact recursive algorithms revealed that the results of the message passing analysis were in fact {\em  exact}: there were neither false positives nor false negatives. Of the 147,892 edges in this network, we correctly identified  all 28,759 bredges in the network. In the case of APs, comparison with exact recursive algorithms revealed just a single false negative: we identified 12,253 out of 62,586 nodes to be APs, missing only one additional AP found by the exact algorithm. Although this demonstrates performance only on a single example, results are certainly encouraging.

The complexity of the algorithm is estimated to scale as $N \ln N$ with system size $N$ for sparse graphs. Here the factor of $N$ accounts for the scaling of the algorithm with the number of edges in a system with finite mean degree, while the $\ln N$ factor accounts for the scaling of the algorithm with the diameter of the network (which scales logarithmically with network size $N$ for small world networks of the type considered in the present paper); the diameter dependence is due to the fact that messages need to be passed (a few times) through the network in order to achieve convergence.

%%%%%%%%%%%%%%%%%%%%%%%%%%%%%%%%%%%%%%%%%%%%%%%%%%%%%%%%%%
\section{Summary and Discussion}
\label{sec:SumDisc}
%%%%%%%%%%%%%%%%%%%%%%%%%%%%%%%%%%%%%%%%%%%%%%%%%%%%%%%%%%
In summary, we demonstrated that the message passing approach to percolation --- apart from its original purpose to compute heterogeneous node-dependent percolation probabilities \cite{KuRog17, KuvM20} --- can also be utilized to evaluate heterogeneous node dependent probabilities of vertices in a complex networks to be APs as well as heterogeneous  edge dependent probabilities of pairs of neighboring vertices in a network to be connected by a bredge.
 
Average probabilities of nodes to be APs and average probabilities of edges to be bredges were recently evaluated for ensembles of networks in the configuration model class in \cite{Tishby+18b} and \cite{Bonneau+20} respectively. In the present paper we looked at the evolution of these probabilities in percolation where a certain fraction of edges is randomly removed from the network, and we were able to go beyond {\em average\/} probabilities. This provides a significant amount of further detail in the analysis. It recognizes and exploits the fact that the probability of a node to be an AP or of an edge to be a bredge, in a node or bond percolation experiment, will depend on the entire environment of the node or edge in question, rather than just on its degree or on the two degrees of the end-nodes of an edge. For the sake of definiteness, we restricted our analysis in the present paper to the case of bond percolation, though it would be straightforward to formulate the theory for site percolation.

We derived a formulation for single instances of large networks and used it to obtain a formulation for ensembles of networks in the configuration model class in the thermodynamic limit. We also obtained closed-form approximations for the large mean degree limit of Erd\H{o}s-R\'enyi (ER) networks which we found to be fairly efficient already for rather moderate values of the mean degree. It is worth emphasizing that solving Eqs.\,\eqref{avnji-eq} for cavity percolation probabilities  is sufficient to obtain node-dependent percolation probabilities $g_i(p)$, using Eq.\,\eqref{avni-eq}, AP probabilities $a_i(p)$, using   Eq.\,\eqref{eq:ai}, and edge dependent bredge probabilities $b_{ij}(p)$, using Eq.\,\eqref{bij}, {\em all in one go}. Alternatively, solving Eq.\,\eqref{pit-gt} for the pdf of cavity percolation probabilities of configuration model networks in the thermodynamic limit is sufficient to obtain limiting pdfs of percolation probabilities, AP probabilities and bredge probabilities from Eqs.\,\eqref{pi-g}, \eqref{pi-a}, and \eqref{pi-b}, respectively, once more {\em all in one go}.

Distributions of AP probabilities and bredge probabilities were evaluated for ER networks as well as scale free networks in the thermodynamic limit; de-convolutions of these distributions according to degree were also obtained. The single instance theory was applied to obtain distributions of AP and bredge probabilities for a real-world network. Finally we also implemented the single instance formulation of the theory {\em prior\/} to averaging over realizations of a percolation experiment as an algorithm to locate APs and bredges in a given network, and for any given realization of a percolation process, and we find that it performs surprisingly well, giving only {\em a single\/} false negative for APs in the Gnutella file-sharing network data --- a network of $N=62,586$ nodes and 147,892 edges, and {\em no\/} errors at all when locating bredges. 

A study of bredges in real world networks \cite{Wu+18} recently revealed that the fraction of bredges in such network was very close to the fraction observe in randomized versions of these networks. In the present paper we have seen that, while this remains true if bonds are randomly removed from either the original network or its randomized version, this is no longer the case for the full distributions of AP and bredge probabilities. 

Articulation points and bredges are exploited in optimized algorithms of network dismantling \cite{Braunstein+PNAS16, Zdeborova+2016, Wandelt+SciRep18}. Dismantling processes typically begin with a decycling stage in which a node is deleted in every cycle of a network. This process transforms the network into a tree or a forest of trees, in which all the nodes of degrees $k \ge 2$ are APs and all edges have turned into bredges. Subsequent removal of further nodes or edges will then break the network into many small components. In this context, the results of the current work and in particular individual $a_i(p)$ and $b_{ij}(p)$ curves  could be useful to devise efficient decycling heuristics that take into account the fraction $1-p$ of edges that the attack is able to delete. In general, a deep attack that aims at dismantling the network completely should initially target nodes/edges with low $a_i(p)$/$b_{ij}(p)$ values in order to achieve decycling, as explained below.

We believe that the quantitative results obtained in this work contain a lot of interesting information about nodes and edges in the network, and could be useful in a broad spectrum of applications. For instance, an interesting aspect of the $b_{ij}(p)$ curves is that they exhibit no crossings, meaning that the order between the $b_{ij}(p)$ curves is preserved in the whole range of $p$'s where they differ. In particular, looking at the $b$ value of all the edges for any intermediate value of $p$ can be used to rank them; the lower the $b_{ij}(p)$ value of an edge $(i,j)$, the more likely it is  that the network would maintain its functionality when the edge in question is removed in the course of a random deletion of of a fraction $1-p$ of its edges. In other words, the $b_{ij}(p)$ curves can be used as a basis for an edge-based centrality measure. We believe that targeting edges with a low $b_{ij}(p)$ score would boost the efficiency of the decycling phase of a deep attack on a network that has a lot of resources, because these edges participate in many cycles. Conversely, targeting high-$b$ edges would result in chipping off fragments from the network even after a short or modest attack, albeit leaving a relatively well connected 2-core behind. 

In contrast to the $b_{ij}(p)$ curves, the $a_i(p)$ curves do exhibit crossings as $p$ is varied and hence cannot provide a $p$-independent way to rank nodes. Still, these curves convey a lot of useful information. For example a node $i$ whose curve $a_i(p)$ starts at $a_i(1)=1$ and connects smoothly to its sub-percolating branch sits on a tree branch or on a finite component. However, if it starts at $a_i(1)=1$ but exhibits a knee before connecting to its sub-percolating branch then it must be a root-AP, namely an AP that sits on the 2-core and glues one or more tree branches to the 2-core. Finally, if the curve starts at $a_i(1)=0$, then the node sits on the 2-core initially, and a deep attack should aim at initially deleting nodes $i$ with the lowest possible $a_i(p)$ values, where $1-p$ is a measure of the effort invested in 
an attack.

A more sophisticated implementation of these ideas could devise an attack with multiple stages, where at each stage only a set of edges or nodes are deleted according to the principles mentioned above. This is followed by a re-assessment of the situation by calculating the updated $a_i(p)$ and $b_{ij}(p)$ curves which could now help decide on the next set of nodes or edges to delete.

Interesting problems to look at in the near future could include generalizing the present analysis to directed networks, which prevail in many technical applications and contexts, and to provide a more systematic study of the single instance message passing algorithm for locating bredges and APs, concerning both its accuracy and the precise scaling of the algorithm with system size. Another aspect one might want to look at is the distribution of the sizes of clusters created by removing APs or bredges from the net, which would require adapting the techniques of \cite{Karrer+14} as used in \cite{KuRog17} to study heterogeneity in percolation to the problem of APs and bredges.

This work was partly supported by the Israel Science Foundation grant no. 1682/18.
%%%%%%%%%%%%%%%%%%%%%%%%%%%%%%%%%%%%%%%%%%%%%%%%%%%%%%%%%%
%\bibliography{../../../MyBib}
\bibliography{hetero-APB}

\begin{thebibliography}{36}
\bibitem[{Bre()}]{Bredge1685}
\bibinfo{note}{{The word `bredge' is an obsolete form of the word `bridge' in
  ancient English; See e.g. English Dictionary, explaining the difficult terms,
  by E. Coles, School-Master and Teacher, printed for Peter Parker at the Leg
  and Star over against the Royal-Exchange in Cornbil (1685)}}.
\end{thebibliography}
%%%%%%%%%%%%%%%%%%%%%%%%%%%%%%%%%%%%%%%%%%%%%%%%%%%%%%%%%%
\end{document}